\documentclass[12pt]{article}

\title{Dynamics of Monatomic Liquids}

\author{Eric D.\ Chisolm and Duane C.\ Wallace \\ T-1, Mail Stop B221 \\ 
        Los Alamos National Laboratory \\ Los Alamos, NM~~87545}

\usepackage{graphics}

\begin{document}

\maketitle

\begin{abstract}

We present a theory of the dynamics of monatomic liquids built on two
basic ideas: (1) The potential surface of the liquid contains three
classes of intersecting nearly-harmonic valleys, one of which (the
``random'' class) vastly outnumbers the others and all whose members
have the same depth and normal mode spectrum; and (2) the motion of
particles in the liquid can be decomposed into oscillations in a
single many-body valley, and nearly instantaneous inter-valley
transitions called {\em transits}.  We review the thermodynamic data
which led to the theory, and we discuss the results of molecular
dynamics (MD) simulations of sodium and Lennard-Jones argon which
support the theory in more detail.  Then we apply the theory to
problems in equilibrium and nonequilibrium statistical mechanics, and
we compare the results to experimental data and MD simulations.  We
also discuss our work in comparison with the QNM and INM research
programs and suggest directions for future research.

\end{abstract}

\section{Introduction}
\label{intro}

Despite a long history of physical studies of the liquid state, no
single theory of liquid dynamics has achieved the nearly universal
acceptance of Boltzmann's theory of gases or Born's theory of lattice
dynamics of crystals.  This shows the extraordinary theoretical
challenge that liquids pose; they enjoy none of the properties that
make either crystals or gases relatively tractable.  A great deal of
effort has been devoted to understanding liquids as hard-sphere
systems, which do model the core repulsion present in real liquids,
but omit the important potential energy effects.  A more realistic
view was given by Frenkel \cite{frenkel1, frenkel2}, who noted that
when a typical crystal melts neither its specific heat, cohesive
properties, nor volume changes greatly, while its diffusion
coefficient increases dramatically; he concluded from this that the
basic motion of particles in a liquid consists of small oscillations
about a set of equilibria, as in a solid, but that these equilibria
are neither symmetrically arranged in space nor unchanging in time.
This highly suggestive picture of a liquid as something like an
amorphous harmonic solid with equilibrium positions that occasionally
move around, allowing for diffusion, has inspired many extensive
programs of research.

For example, after Stillinger and Weber's computer simulations
\cite{stillweb1, stillweb2} revealed the existence in the liquid of
mechanically stable arrangements of particles, called inherent
structures, with a wide range of energies, several workers have
developed the idea that the liquid moves in a ``rugged potential
energy landscape'' with a wide distribution of structural potential
energies separated by barriers having a wide distribution of heights
\cite{still, rabgezber1}.  These ideas have influenced the development
of both the quenched normal mode, or QNM \cite{rabgezber1, 1qnm,
rabgezber2} and instantaneous normal mode, or INM \cite{keyes1,
keyes2, stratt1, stratt2, stratt3, benlaird, wutsay} schools of
thought.

Here we will review another line of inquiry which has been pursued for
the last five years or so \cite{wall1, wall2, wall3, wall4, wall5,
wall6, wall7, wall8, wall9} and which differs from the others in that it
begins by focusing on a restricted class of liquids (see below), and
it proposes that they move in a significantly simpler potential
landscape.  

This work is concerned exclusively with monatomic liquids, meaning
elemental liquids which do not exhibit molecular bonding.  Monatomic
liquids include all elemental liquid metals and the rare gas liquids,
but not molecular liquids such as N$_2$ and O$_2$, and not polyatomic
systems such as alkali halides or water.  Molecular liquids have
translational, rotational, and internal vibrational degrees of
freedom, while monatomic liquids have only translational motion, and
the potential energy surface for monatomic liquids is presumably the
simplest of all liquid potential landscapes.  Our strategy is to
develop a thorough understanding of this hopefully simplest case, and
then to apply the insights gained there to more complex liquid
systems.

In Section \ref{picture}, we describe the thermodynamic data which led
us to a specific picture of liquid dynamics, and we describe the
picture itself in some detail.  As will be clear, the thermodynamic
data are consistent with the picture, but they do not lead to it
uniquely; thus additional support is called for.  In Section
\ref{verify} we review the results of molecular dynamics (MD) studies
of two particular liquids, sodium and Lennard-Jones argon, which
support many of our claims in far more detail.  Then we apply the
picture in Sections \ref{eq} and \ref{noneq} to problems in
equilibrium and nonequilibrium statistical mechanics, and we compare
the results to experimental data and MD simulations.  In Section
\ref{outlook}, we briefly review the picture, compare it with other
research programs, consider the current status of its verification,
discuss further problems to which it may be applied, and describe the
role we believe it fills in the continuing effort to develop a
comprehensive theory of the dynamics of liquids.

\section{The picture}
\label{picture}

\subsection{Thermodynamic data}

Initial support for our picture comes from an analysis of two types of
thermodynamic data: The constant-volume specific heat $C_V$ at the
melting point of various monatomic liquids, and the entropy of melting
of these elements.

\subsubsection{Specific heat}

\begin{figure}
\includegraphics*[100,190][700,700]{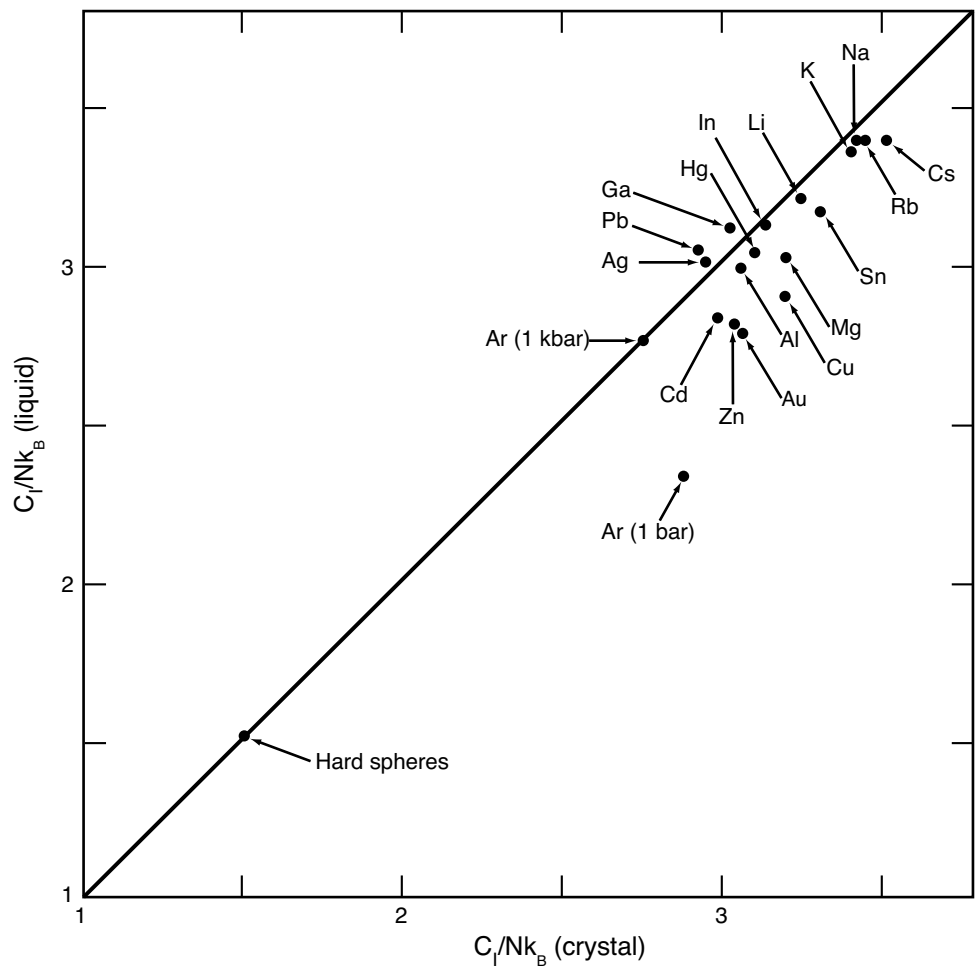}
\caption{Ion-motional specific heat for 18 elements in both liquid and 
         crystal phases at melt.  Adapted from \cite{wall2}.}
\label{CV}
\end{figure}

The experimentally determined specific heats at constant pressure
$C_P$ for the elements have been compiled by Hultgren et al.\
\cite{hultgren} and Chase et al.\ \cite{JANAF} for both crystal and
liquid phases at the melting point; these can be corrected in the
standard way to determine $C_V$.  $C_V$ is composed of the
contributions $C_I$ from the motion of the ions and $C_E$ from the
excitation of the valence electrons,
\begin{equation} C_V = C_I + C_E, \end{equation}
and for the nearly-free-electron elements the electronic contribution 
is given accurately by
\begin{equation} 
C_E = \frac{1}{3} \pi^2 N k_B^2\,T\,n(\epsilon_F), 
\end{equation}
where $n(\epsilon_F)$ is the electron density of states per atom at
the Fermi energy $\epsilon_F$.  Thus Wallace \cite{wall1} chose to
study the nearly-free-electron elements, for which $C_I$ can be
accurately determined.  He took $n(\epsilon_F)$ from band structure
calculations when possible \cite{band1, band2} and from free-electron
theory otherwise; then he subtracted out the electronic contribution
to $C_V$, and the resulting ionic contributions for both the crystal
and liquid phases are shown in Figure \ref{CV}, which is adapted from
Figure 1 of \cite{wall2}.  The quantities predicted by hard-sphere
theory are shown for comparison.
Notice that all elements cluster around $C_I = 3Nk_B$ in both
phases.  (The exception is argon at 1 bar, which is known to be rather
gaslike, but at pressures approaching 1 kbar its behavior more closely
resembles that of the other liquids; thus, we will henceforth consider
only compressed argon.)  It is known that any of the crystals may
be modeled very accurately as a set of $3N$ harmonic oscillators, thus
accounting for their specific heats; this is the starting point of
lattice dynamics.  That the liquids at melt have nearly the same
values for $C_I$ suggests that they too behave as harmonic
oscillators.  The departures from harmonicity for both phases lie
outside the experimental errors; anharmonic effects on the $C_I$ of 
liquids will be discussed in Sections \ref{eq} and \ref{outlook}.

\subsubsection{Entropy of melting}
\label{entmelt}

A study of the entropy of melting at constant density (but not
constant pressure) of a large number of monatomic liquids led Wallace
\cite{melt, thermeval} to suggest that the elements can be separated
into two classes: the ``normal melting elements,'' for which $\Delta S
= (0.80 \pm 0.10)N k_B$, and the ``anomalous melting elements,'' for
all of which $\Delta S$ lies far above the range of the normal
melters.  The entropy of melting results are shown in Table \ref{DS},
which is adapted from Table III of \cite{thermeval}.  The first column
is the set of normal melters used to calculate the number given above;
the second is a set of transition metals, which reasonably may be
considered normal melters (but not anomalous melters) given the larger
errors in their $\Delta S$ data.  The six anomalous melters are in the
final column.  The electronic structures of the normal melters do not
change greatly upon melting, while for the anomalous melters the
structure change is noticeable (semimetal crystal to metal liquid,
semiconductor crystal to metal liquid, etc.), and this change is
presumably responsible for the excess contribution to $\Delta S$.
\begin{table}
\begin{tabular}{lclclc} \hline\hline
Element & $\Delta S / N k_B$ & Element & $\Delta S / N k_B$ & 
                                     Element & $\Delta S / N k_B$ \\ \hline
Li & 0.75 & V  & 0.90  & Sn & 1.48 \\
Na & 0.73 & Nb & 0.97  & Ga & 2.37 \\
K  & 0.73 & Ta & 1.1   & Sb & 2.68 \\
Rb & 0.73 & Cr & (0.9) & Bi & 2.62 \\
Cs & 0.73 & Mo & (1.2) & Si & 3.77 \\
Ba & 0.90 & W  & (1.1) & Ge & 3.85 \\
Fe & 0.68 & Pd & 0.74  &    &      \\
Al & 0.88 & Pt & 0.79  &    &      \\
Pb & 0.68 & Ti & 0.70  &    &      \\
Cu & 0.86 & Zr & 0.93  &    &      \\
Ag & 0.73 &    &       &    &      \\
Au & 0.64 &    &       &    &      \\
Ni & 0.88 &    &       &    &      \\
Mg & 0.96 &    &       &    &      \\
Zn & 0.97 &    &       &    &      \\
Cd & 0.93 &    &       &    &      \\
In & 0.76 &    &       &    &      \\
Hg & 0.90 &    &       &    &      \\ \hline\hline
\end{tabular}
\caption{Entropy of melting at constant density for 34 elements.  The 
         normal melting elements are in the first two columns, and the 
         anomalous melting elements are in the third column.  Data in  
         parentheses are less reliable.  Adapted from \cite{thermeval}.}
\label{DS}
\end{table}
Considering only the normal melting elements for a moment, this change
in entropy upon melting is consistent with a scenario in which the
system, which had previously been confined to a single crystalline
potential valley, upon melting suddenly finds itself able to move over
a space of approximately $w^N$ valleys where $\ln w = 0.8$; the
entropy increase is due to the greater size of the available
configuration space.  (Strictly speaking, this is true only if certain
other restrictions are satisfied; see Subsection \ref{exp} for a more
extensive discussion.)  We hypothesize that this change occurs for all
melters, normal and anomalous, and that additional factors deriving
from the change in the electronic structure account for the difference
between the two cases.  We return to the anomalous melters in Section
\ref{outlook}.

\subsection{Details of the picture}
\label{details}

The data considered so far suggest that particles in a liquid move in
a potential landscape dominated by harmonic valleys.  We have refined
this observation into a more precise picture of both the motion of the
particles and the nature of the potential energy surface in which they
move.

\subsubsection{The motion}

We hypothesize that the motion of the system may be decomposed into
\mbox{two} distinct types: Oscillation in a single nearly harmonic
many-body valley, and nearly instantaneous transitions between valleys
which we call \mbox{\em transits}.  That the valleys are nearly
harmonic, and that the transits are nearly instantaneous, are both
suggested by the $C_I$ data, since $C_I$ in all cases is quite close
to the value expected for equilibrium motion in a single harmonic
valley.  In fact, any significant departure from this behavior should
show itself clearly in the $C_I$ data, so we believe that this part of
the picture is very solidly supported by experiment.  (Higher order
corrections, to account for the \mbox{small} deviations of $C_I$ from
precisely $3Nk_B$, are considered in Subsection \ref{thermo} and
discussed in more detail in Section \ref{outlook}.)  We also expect
transits to involve only a few particles in the system at a time,
because transits perform a function in liquids analogous to that
performed by collisions in a Boltzmann gas: They drive the system
irreversibly toward equilibrium, and once there, they maintain
equilibrium by constantly opposing fluctuations.  Mechanisms of
equilibration operate on a local level, since any small region can
equilibrate independently of the rest (except for equilibria involving
macroscopic coherent quantum states, not considered here), so we
expect transits to operate locally.  Unlike gases, though, in which
collisions almost always involve only two particles at a time, in
liquids slightly larger groups of particles can undergo cooperative
motion, since they are sufficiently close together that interparticle
potentials are always significant, so a single transit could involve
as many as tens of particles.

\subsubsection{Types of potential valleys}

If the valleys in the many-body potential surface are nearly harmonic,
then each is characterized by its structure potential $\Phi_0$,
defined to be the value of the system's potential energy at the bottom
of the valley, and its density of normal mode frequencies $g(\omega)$.
We also hypothesize that the valleys may be divided into three
classes: crystalline, symmetric, and random.

(1) A {\em crystalline} valley is occupied when the system is in
one of its crystalline phases.  These valleys are very few in number,
and since the crystalline phases are the most stable at low
temperatures, they also have the lowest value for their structure
potential $\Phi_0$.  Due to their very small number,
the crystalline valleys make a negligible contribution to the
statistical mechanics of the liquid.

(2) The {\em symmetric} valleys correspond to more disordered
configurations that still retain some remnant of crystalline symmetry.
This group includes a large variety of polycrystalline and
microcrystalline types, as well as the states of carbon realized
experimentally by McKenzie, Muller, and Pailthorpe \cite{mcken} which
differ from the perfect diamond by irregularly distorted bond lengths
and angles.  These valleys are more shallow than the crystalline ones,
and their structure potentials $\Phi_0$ are expected to cover a wide
range of values due to their large variety of symmetry properties.
Also because of their widely varying symmetries, we expect the normal
mode spectrum $g(\omega)$ to vary substantially from valley to valley.

(3) Finally, the {\em random} valleys are occupied when the system
retains no remnant crystalline symmetries.  Since their configurations
suffer no symmetry restrictions, these valleys should greatly
outnumber both the crystalline and symmetric valleys; in fact, we
hypothesize that almost all of the $w^N$ valleys available to the
liquid are random, so the random valleys dominate the statistical
mechanics of the liquid.  Further, since the random valleys have no
symmetry properties that allow them to be distinguished from one
another, we expect that in the large-$N$ limit all random valleys
should have the same structure potential $\Phi_0$ and normal mode
spectrum $g(\omega)$, in stark contrast to both the crystalline and
symmetric valleys.  (In the examples we have studied, $\Phi_0$ for the
random valleys always lies above the $\Phi_0$ values for all of the
symmetric valleys, but we see no reason for this to be true over the
entire potential surface.)

The hypothesis that the vast majority of valleys available to a
monatomic liquid have the same depth and vibrational spectrum is a
distinctive part of our approach, and it has extraordinary
consequences for the statistical mechanics of the liquid; however, it
is clear that the data considered to this point lend that idea scant
support.  Thus further studies were conducted to test the validity of
this picture for specific monatomic liquids; these studies are
discussed next.

\section{Verifying the picture}
\label{verify}

Our picture of monatomic liquids consists of two sets of hypotheses:
Those concerning the motion of the system, particularly that transits
occur rapidly and involve only a few particles; and those concerning
the potential energy surface and the classification of valleys into
three types.  We consider tests of each set of hypotheses in turn.

\subsection{Transits}
\label{transits}

To investigate the properties of transits, we conducted computer
simulations of two liquids: sodium and Lennard-Jones argon.

\subsubsection{Sodium}

Our simulation of an $N$-atom sodium system is described in detail in
\cite{wall4}.  The particles interact through a potential of the
general form \cite{har, wall}
\begin{equation}
\Phi(\{\mbox{\boldmath $r$}_K\}) = \Omega(V) + \frac{1}{2} \sum_{K,L}
              \phi(|\mbox{\boldmath $r$}_K - \mbox{\boldmath $r$}_L|; V),
\label{pot}
\end{equation}
where the strongly negative $\Omega(V)$ is responsible for metallic
binding and the effective ion-ion potential $\phi(r; V)$ is given by
pseudopotential theory \cite{pseudopot}.  This pair potential is shown
in Figure 1 of \cite{wall4}; it is multiplied by a damping factor to
remove long-range Friedel oscillations, and this is the only
significant effect of the factor on the potential.  After being
calibrated to the bulk properties of crystalline sodium at $0$ K, the
full potential in Equation (\ref{pot}) has been shown to reproduce
with remarkable accuracy several known properties of metallic sodium,
such as the phonon frequency spectrum and the melting temperature as a
function of pressure.  In our simulations, the volume per atom $V_A =
V/N$ was fixed at $278\,a_0^3$, where $a_0$ is the Bohr radius; this
is the density of liquid sodium at melt when the pressure is 1 bar and
the melting temperature is $371$ K.  Since $V$ is held constant in our
MD calculations, we chose to set $\Omega(V)$ to zero.  The rms
vibrational frequency of a typical many-body valley in this potential
is $1.56 \times 10^{13}$ s$^{-1}$.  (See Subsection \ref{structure}
for more on the structure of potential valleys in sodium.)
Calculations were performed using the Verlet algorithm \cite{verlet}
for a system in a cubical box with periodic boundary conditions; the
natural time scale of the system is $t^* = \sqrt{2Ma_0^3/e^2}$, where
$M$ is the atomic mass of sodium, or $t^* = 7.00 \times 10^{-15}$ s.
(The mean vibrational period in a typical potential valley is $\tau =
57.45\,t^*$.)  The two parameters which varied between runs were the
number of particles $N$ and the MD time step $\delta t$, which was
always taken to be some fraction of $t^*$.  We will refer frequently
to MD studies of sodium through the rest of the paper, and each time
we will indicate the values of each of these parameters.

\begin{figure}
\includegraphics{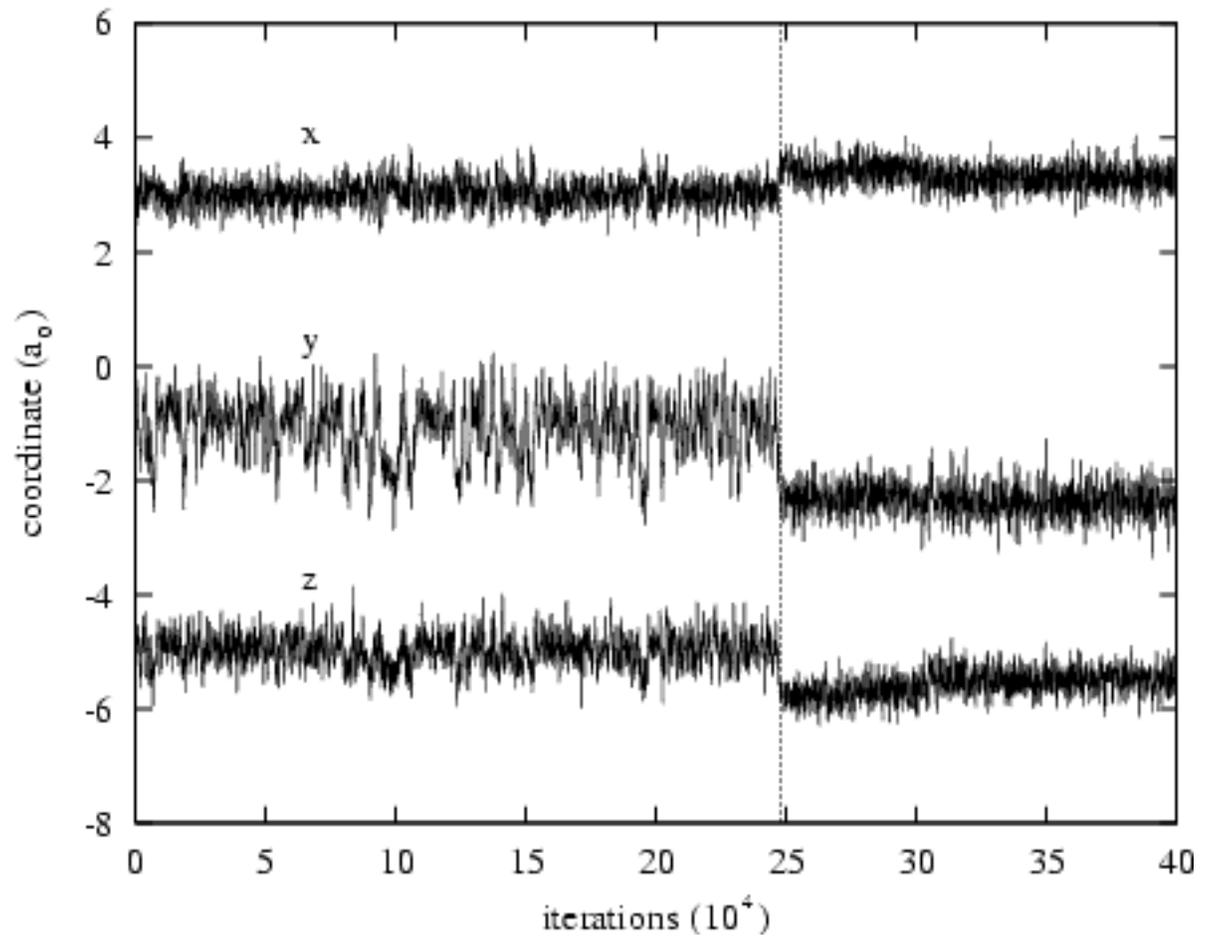}
\caption{The coordinates of one particle in an 11-particle transit in sodium 
         at $30$ K.  Adapted from \cite{wall9}.}
\label{natran}
\end{figure}

We searched for transits in an $N = 500$ system where the time step was
set to $\delta t = 0.2\,t^*$.  We cooled the system to a sufficiently
low temperature that once it had equilibrated it remained in a single
valley, as could be verified from its mean-squared displacement.  We
then raised the temperature by very small increments, each time
allowing the system to equilibrate again, until transits began to
occur at $T = 30$ K.  (The details of our method of searching for
transits may be found in \cite{wall9}.)  The $x, y$, and $z$
coordinates of a particle in a typical transit are shown in Figure
\ref{natran}.
Our general observations in sodium are as follows \cite{wall9}: Every
particle in the system either oscillated for the entire run around a
single location, or it executed a transit of the general type seen in
Figure \ref{natran}, where the particle oscillated in a single region
of space for some time, abruptly moved to a new region, and continued
to oscillate in the new region.  Typically small groups of particles
transited simultaneously, and many more particles would execute
smaller shifts in their equilibrium positions during a small window in
time around the transit.  Further, it was not uncommon for a single
particle to participate in two or three transits, well separated from
one another in time.  The average shift in the equilibrium position of
a particle involved in a transit was $1.75\,a_0$ (about one quarter of
the nearest neighbor distance of $7a_0$); the average duration in time
of any transit was $\tau$, and this includes the time taken by
precursors and postcursors to some of the transits (described below in
the discussion of argon).  Thus our general picture of transits as
abrupt transitions between equilibrium positions of a small group of
particles is supported in this system.

\subsubsection{Lennard-Jones argon}

Our simulation of argon consists of $N$ particles interacting through a 
Lennard-Jones pair potential
\begin{equation}
\phi(r) = 4\epsilon\left[\left(\frac{\sigma}{r}\right)^{12} - 
                    \left(\frac{\sigma}{r}\right)^6\right]
\end{equation}
where $\epsilon = 1.65 \times 10^{-21}$\,J (with equivalent
temperature $119.8$ K) and $\sigma = 3.405$ \AA.  The density was set
to $0.9522$ particles/$\sigma^3$, or $1.600$ g/cm$^3$.  (The density
of liquid argon at melt at 1 bar is $1.414$ g/cm$^3$.)  The rms
vibrational frequency of a typical many-body valley in this potential
is $6.88 \times 10^{12}$ s$^{-1}$.  (See Subsection \ref{structure}
for more on the structure of potential valleys in argon.)
Calculations were performed again using the Verlet algorithm for a box
with periodic boundary conditions; the natural time scale of this
system is $t^* = \sqrt{M\sigma^2/\epsilon}$, where $M$ is the atomic
mass of sodium, or $t^* = 2.16 \times 10^{-12}$ s.  (The mean
vibrational period in a typical potential valley is $\tau =
0.424\,t^*$.)  The time step in all MD calculations was $\delta t =
0.001\,t^*$; the only parameter varied between argon runs was $N$.

\begin{figure}
\includegraphics{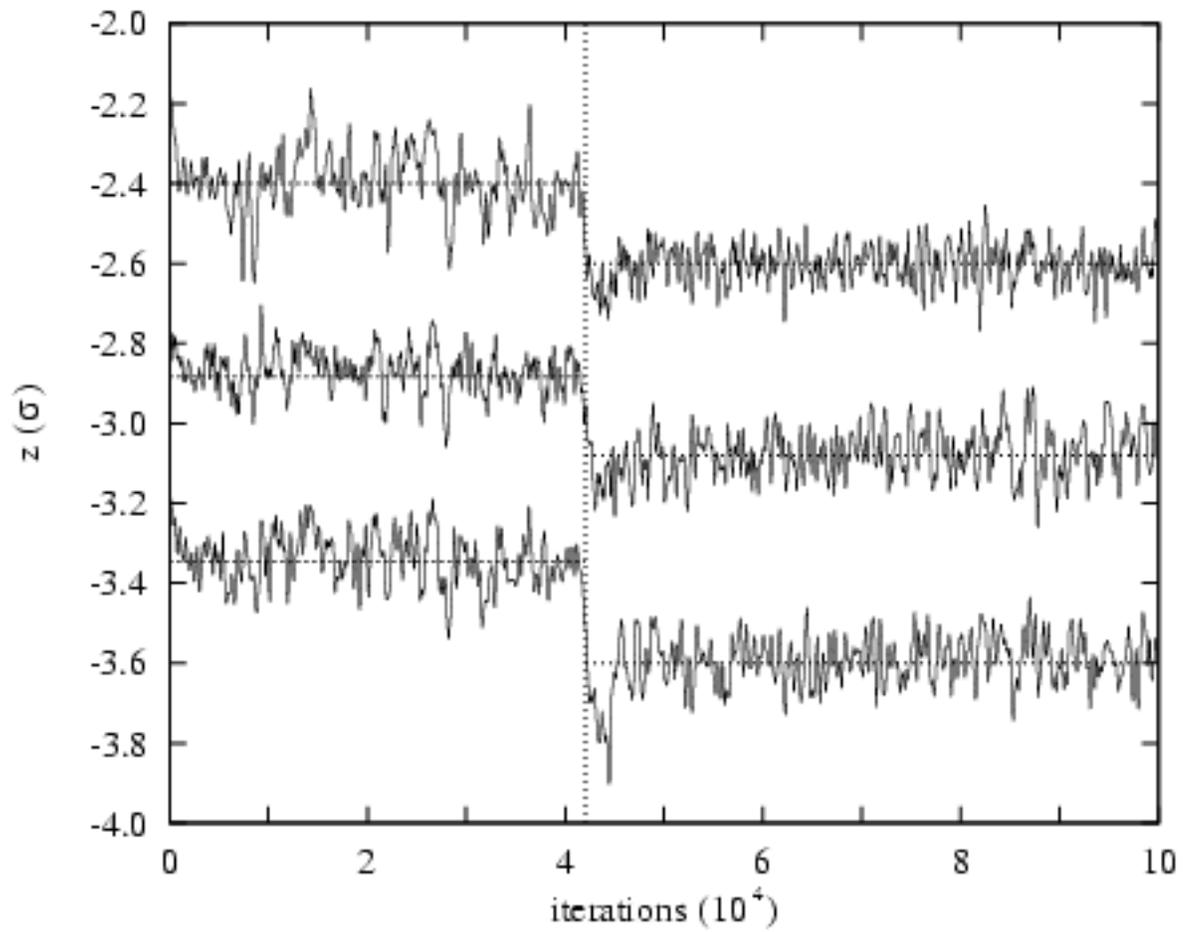}
\caption{The $z$ coordinates of three particles from an 8-particle transit 
         in Lennard-Jones argon at $17.1$ K.  Adapted from \cite{wall9}.}
\label{artran1}
\end{figure}

\begin{figure}
\includegraphics{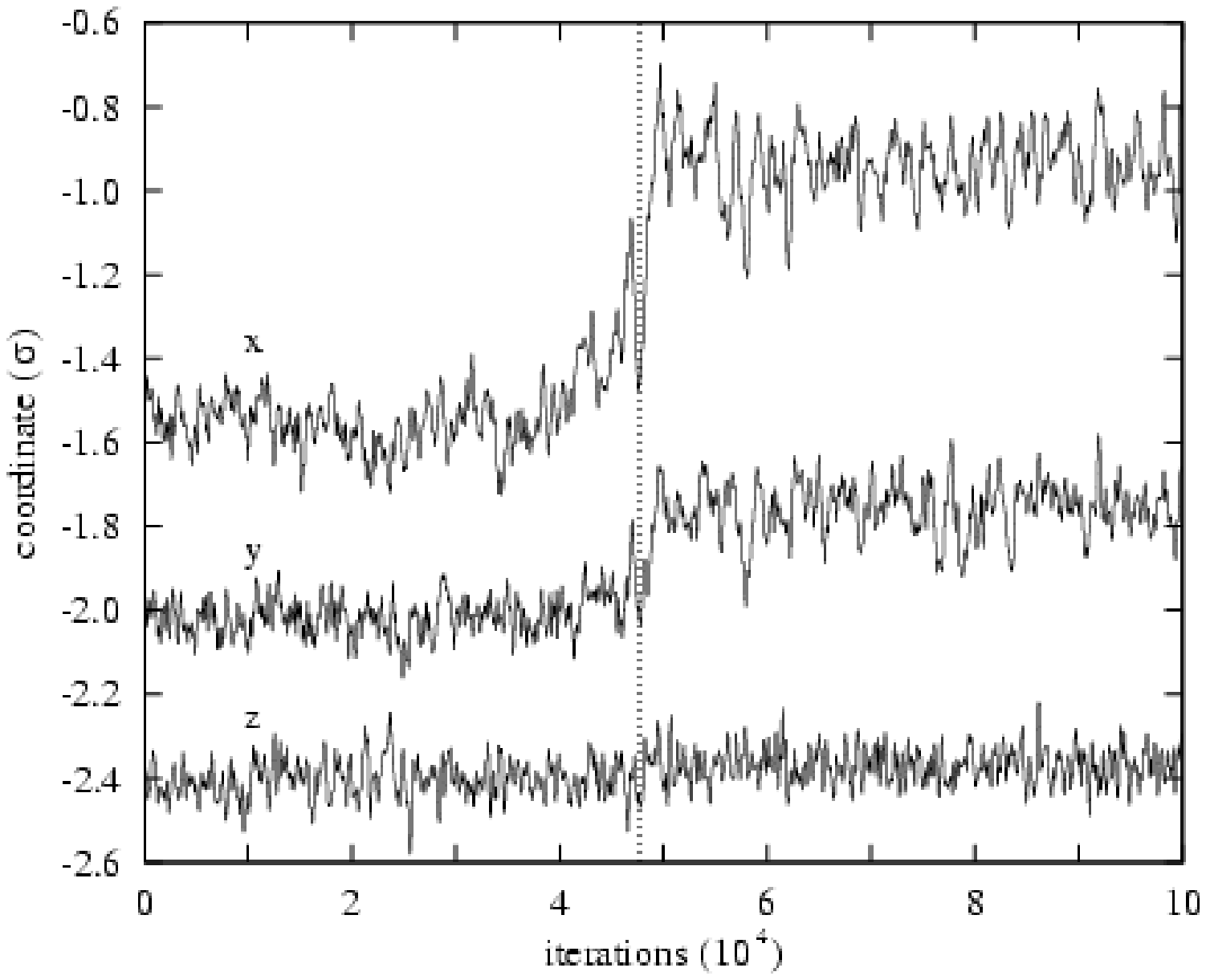}
\caption{The coordinates of one of three particles involved in a transit 
         in Lennard-Jones argon at $17.1$ K.  Note the precursor, which 
         is particularly visible in the $x$ coordinate.  Adapted from 
         \cite{wall9}.}
\label{artran2}
\end{figure}

We searched for transits in an $N = 500$ system using the same
technique that was used for sodium; we found transits at $17.1$ K.
The $z$ coordinates of three particles involved in an 8-particle
transit are shown in Figure \ref{artran1}.
The horizontal dotted lines indicate the equilibrium positions of the
particles before and after the transit; the vertical line indicates
the transit time.  All of the general observations made above about
transits in sodium also hold here \cite{wall9}: The type of motion
seen in Figure \ref{artran1} is typical, usually small groups of
particles transited simultaneously, and individual particles sometimes
participated in multiple distinct transits.  The average shift in the
equilibrium position of a particle involved in a transit was
$0.44\,\sigma$ (about four tenths of the nearest neighbor distance of
$1.095\,\sigma$); the average duration in time of any transit was
again $\tau$, and again this includes the time taken by precursors and
postcursors to some of the transits.  By a precursor or postcursor, we
mean a slow drift by a single particle into a new equilibrium position
either before or after a multiple-particle transit; a typical
precursor is shown in Figure \ref{artran2}, part of the record of a
3-particle transit that occurred roughly $13\,\tau$ after the transit
shown in Figure \ref{artran1}.
Every drift of the type seen in Figure \ref{artran2} that we found
occurred in connection with a transit, so we believe that precursors
and postcursors are part of the transit process.  They are the primary
reason that the duration of a typical transit in either system is as
high as $\tau$; many transits are of essentially zero duration when
one neglects these effects, and most transits exhibit no precursors or
postcursors and so are genuinely \mbox{nearly} instantaneous.  Thus we
conclude that our basic picture of the motion of particles in a liquid
as a combination of oscillations and transits is verified in these
cases.  The precursors and postcursors are still of some interest,
however, and we will comment on them again in Section \ref{outlook}.

\subsection{Structure of the many-body potential landscape}
\label{structure}

\subsubsection{Sodium}

\begin{figure}
\includegraphics[50,50][575,575]{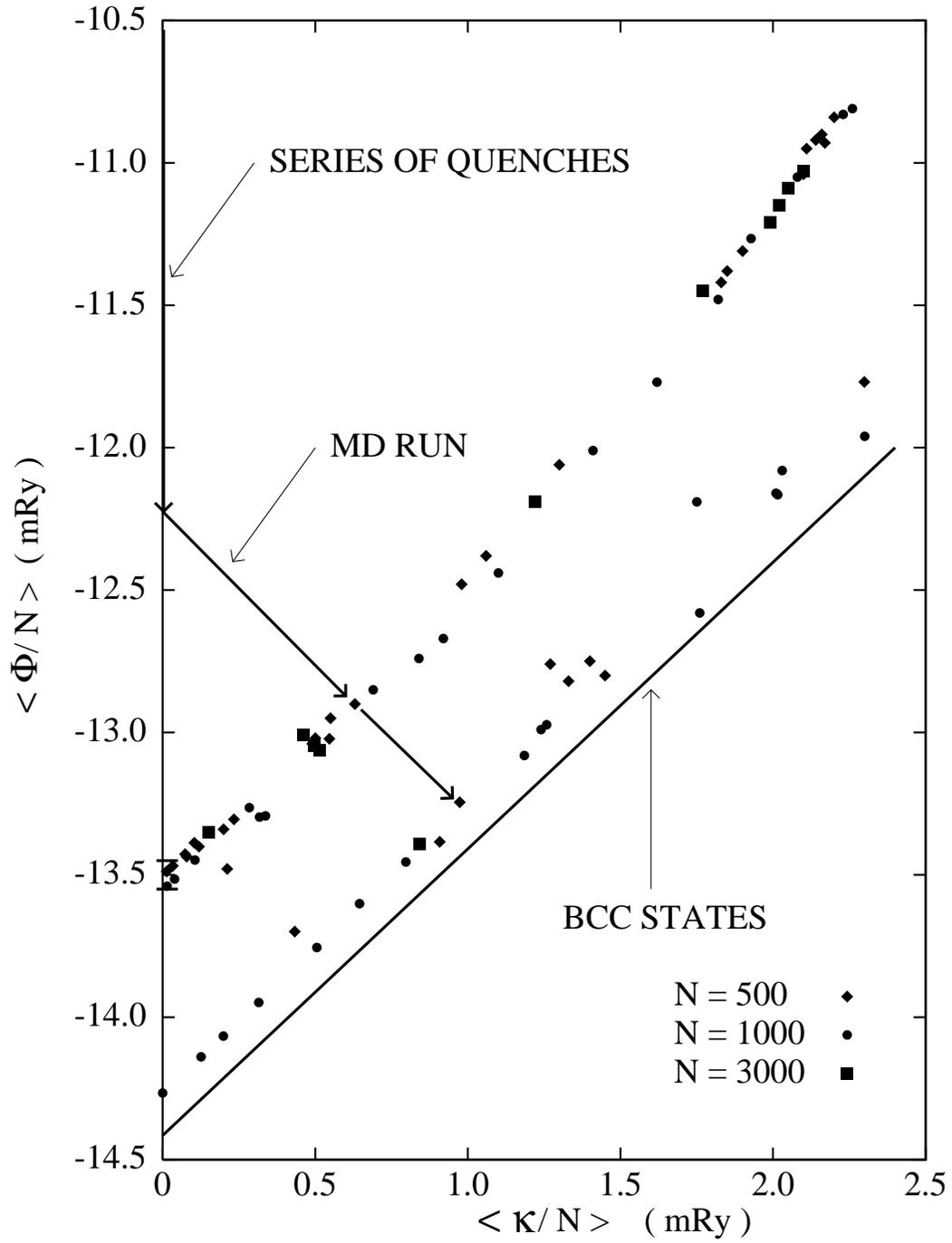}
\caption{$\langle \Phi / N \rangle$ versus $\langle \mbox{$\cal{K}$} / N 
         \rangle$ for several equilibrium states in sodium.  From 
         \cite{wall4}.}
\label{pvsk}
\end{figure}

Wallace and Clements \cite{wall4, wall5} conducted an exhaustive study
of the many-body potential underlying the sodium simulations in order
to test the validity of the three-fold classification of valleys
proposed above.  They generated a large number of supercooled
equilibrium states of systems with $N$ = 500, 1000, and 3000 and
cataloged properties such as their energies and pair distribution
functions.  They made the following observations about the states:

(1) A graph of time-averaged potential energy per particle $\langle
\Phi / N \rangle$ versus time-averaged kinetic energy per particle
$\langle \mbox{$\cal{K}$} / N \rangle$ for the equilibrium states is
shown in Figure \ref{pvsk}.
The melting temperature $T = 371$ K corresponds to $\langle
\mbox{$\cal{K}$} / N \rangle = 3.53$ mRy, so all of the states in the
figure are supercooled, as claimed.  This figure also shows the curve
occupied by the bcc crystal states and the path followed by a typical
MD run used to generate the states: Several quenches keep the kinetic
energy at zero while the system moves down the path of steepest
descent on the potential energy surface, so its potential energy
continues to decrease; and when the quenches end the system
equilibrates under the condition that $\Phi + \cal{K}$ remains
constant, so the system moves down the 45$^{\circ}$ line on the graph.
Notice that the states separate cleanly into two distinct groups.
Each group of states lies approximately along a line with unit slope,
as predicted by the equipartition theorem if the states are moving in
harmonic valleys, although the lower group shows considerable scatter
and the slope of the upper line increases at higher temperatures.
Thus we tentatively suggest that for each $N$ the system is moving in
a landscape of approximately harmonic valleys, but we also see
significant anharmonic effects to which we will return later.

(2) The system almost always quenched into one of the states from the
upper group first; if the temperature was between approximately $35$ K
and $200$ K, it would remain in such a state for several thousand time
steps (long enough to compute equilibrium data) before settling
spontaneously into one of the states in the lower group.  It would
remain in this state for as long as the MD run proceeded.

(3) The states in the upper group lie along the same curve as the
equilibrium states of the liquid, while the states from the lower group 
appear to be bounded in energy by the limits of the graph.

\begin{figure}
\includegraphics[100,50][300,400]{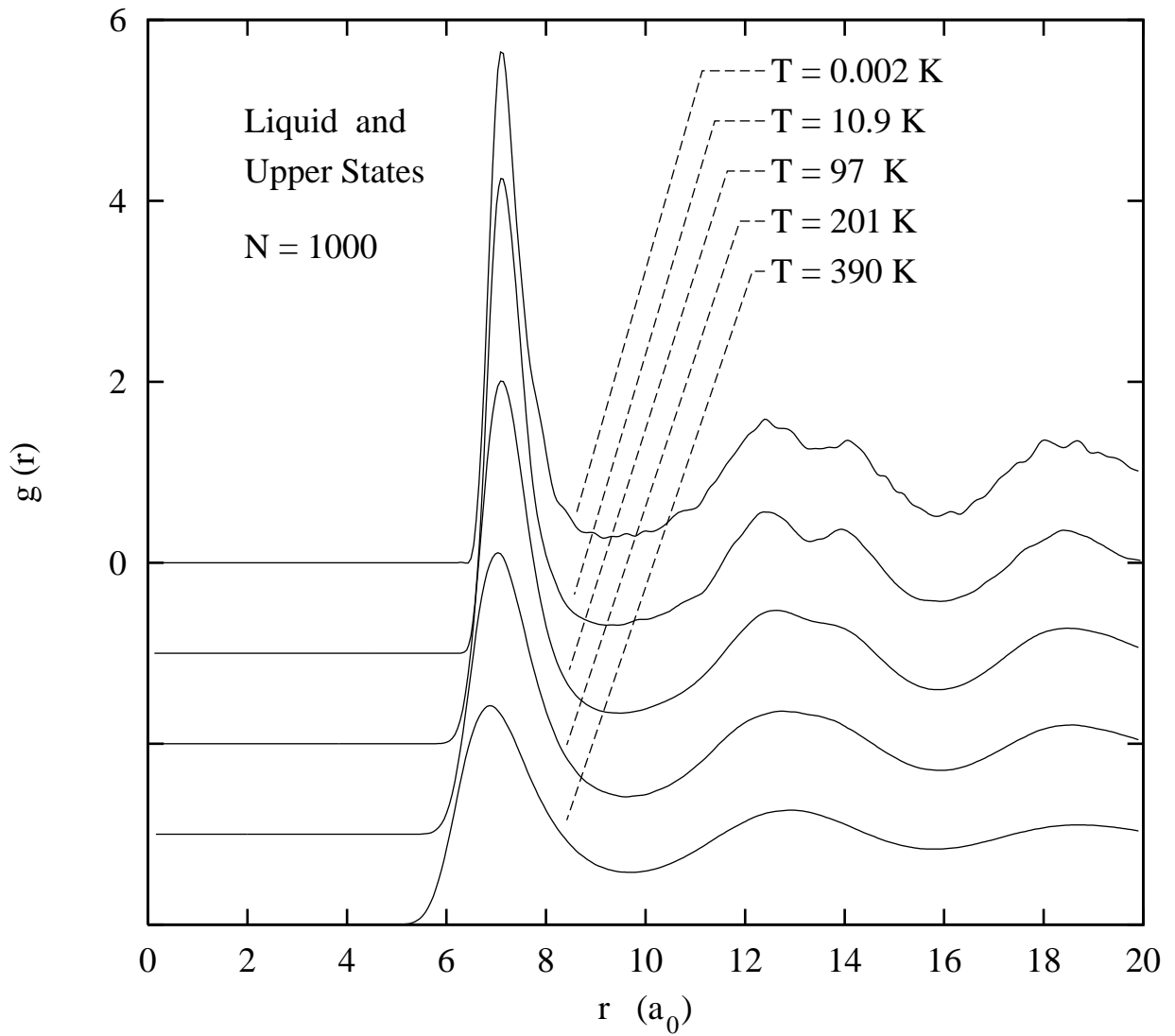}
\caption{The pair distribution function for the liquid at $390$ K and 
         pair distribution functions for various states from the upper 
         group at temperatures from $0.002$ K to $201$ K.  Notice how 
         the states continuously evolve into the liquid state with 
         increasing temperature.  Adapted from \cite{wall5}.}
\label{toliq}
\end{figure}

(4) As $T$ is increased, the graph of the pair distribution function
$g(r)$ for the states from the upper group smoothly evolves into
$g(r)$ for the liquid state, as shown in Figure \ref{toliq}.

(5) For a state from the lower group at temperatures above $100$ K,
$g(r)$ exhibits a split second peak, with the first subpeak lower than 
the second.

\begin{figure}
\includegraphics[100,50][300,400]{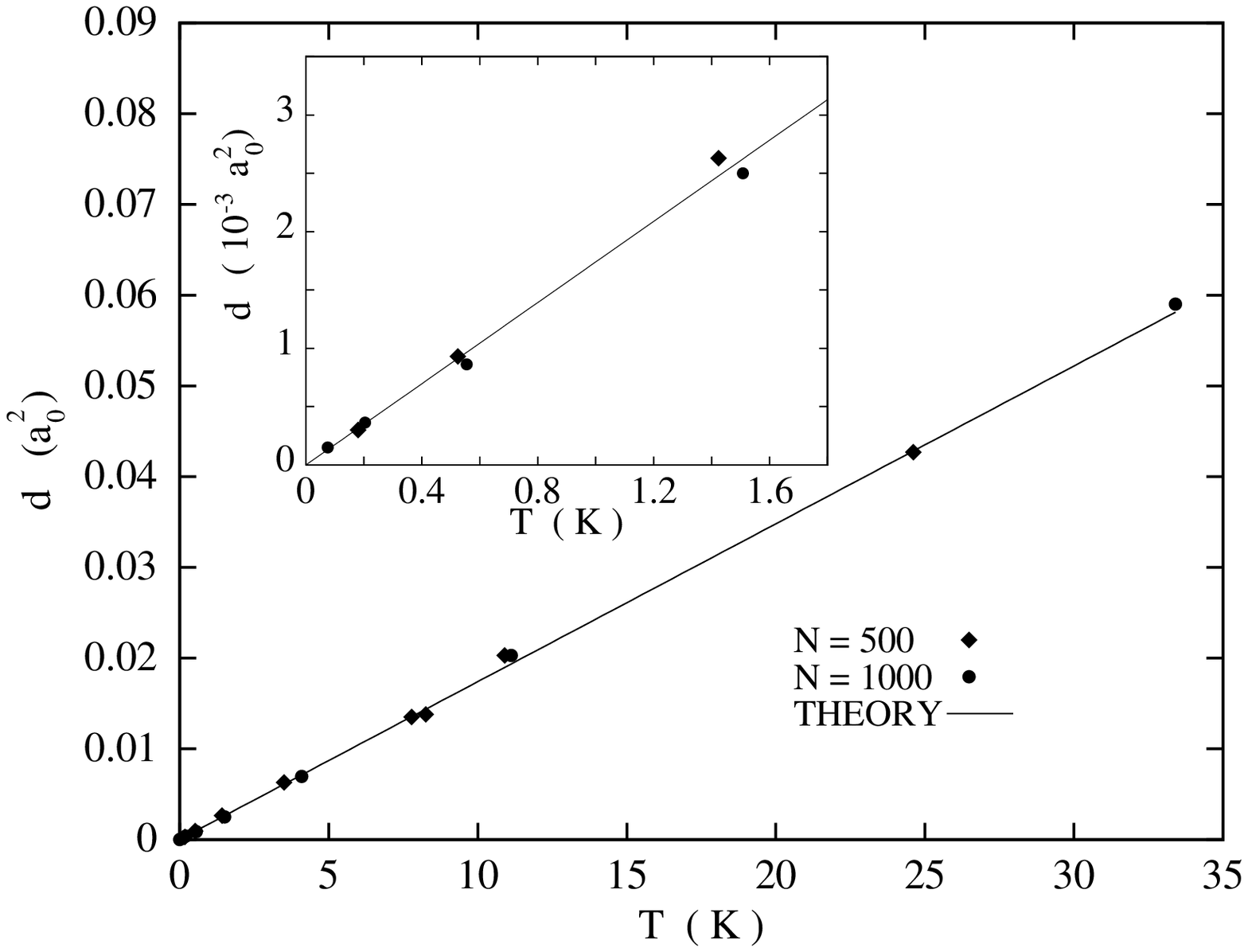}
\caption{$d$ versus $T$ for several confined states in the upper group 
         compared with the harmonic prediction.  The inset figure shows the 
         low-temperature data in more detail.  Adapted from \cite{wall4}.}
\label{DvsTrand}
\end{figure}

(6)  By observing the mean-squared displacement of each state, defined by
\begin{equation}
d(t) = \frac{1}{6N} \sum_K \left[\mbox{\boldmath $r$}_K(t) - 
       \mbox{\boldmath $r$}_K(0)\right]^2,
\end{equation}
Wallace and Clements found that low-temperature states from both
groups were confined to individual valleys of the potential surface.
Let $d$ be the time average of $d(t)$, or $d = \langle d(t)
\rangle_t$.  Then for a system in equilibrium in a single many-body
harmonic valley,
\begin{equation}
d = \frac{3 \hbar^2 T}{M k_B \Theta_{-2}^2},
\label{harmD}
\end{equation}
where $\Theta_{-2}$ (defined below) is one of the principal moments of
the valley's frequency distribution.  (For a derivation, including
some subtleties involving the omission of zero-frequency modes
corresponding to center of mass motion, see \cite{wall4}.)  Thus, if
these states are confined in harmonic valleys, $d$ should be a linear
function of $T$.  As we will note below, all of the valleys occupied
by confined states in the upper group have the same frequency
distribution, and thus the same $\Theta_{-2}$; Figure \ref{DvsTrand}
shows $d$ for several confined states in the upper group compared with 
Equation (\ref{harmD}) using the common value of $\Theta_{-2}$.
The superb agreement further suggests that the valleys in which these
states are trapped are in fact harmonic.
Figure 13 of \cite{wall4} shows that the same relation holds for
several states from the lower group, all of which are confined to the
same valley; however, $\Theta_{-2}$ is different for different valleys
occupied by lower states (see below), so states from different valleys
fit curves with different slopes.

\begin{figure}
\includegraphics[100,50][300,400]{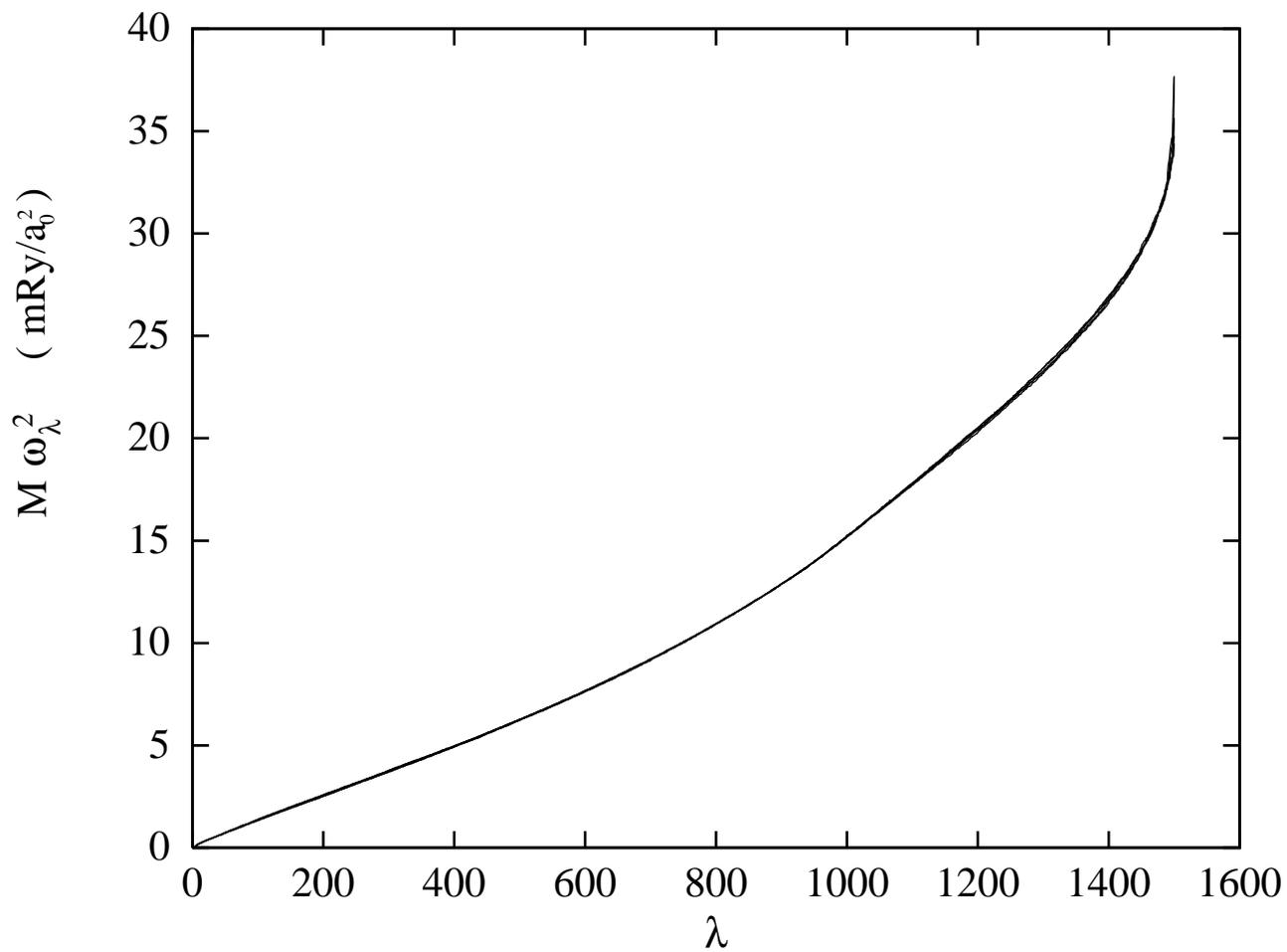}
\caption{The normal mode spectra for five different valleys occupied by 
         upper equilibrium states.  Eigenvalues are given as $M 
         \omega_{\lambda}^2$ where $M$ is the atomic mass of sodium and 
         $\lambda$ is a label that counts the eigenvalues.  From 
         \cite{wall4}.}
\label{freq}
\end{figure}

Next, Wallace and Clements studied the actual many-body potential
valleys occupied by the confined states, determining properties such
as each valley's depth $\Phi_0$ and vibrational frequency spectrum
$g(\omega)$.  They made the following observations about the valleys:

(1) The depths of the valleys occupied by the upper states all lie in a 
very narrow range,
\begin{equation}
\Phi_0 / N = -0.01352 \pm 0.00002\ {\rm Ry/particle,}
\end{equation}
and they all have virtually the same normal mode frequency spectrum
independent of the valley or even $N$.  The normal mode spectra for
five such valleys are shown in Figure \ref{freq}.
Since the normal mode frequencies for different valleys are so
similar, it comes as no surprise that the three principal moments of
the frequency distribution, $\Theta_{-2}, \Theta_0,$ and $\Theta_2$,
defined by 
\begin{eqnarray}
k_B \Theta_{-2} & = & \left[\frac{1}{3} \left\langle (\hbar \omega)^{-2} 
                    \right\rangle \right]^{-1/2} \nonumber \\
\ln k_B \Theta_{0} & = & \langle \ln \hbar \omega \rangle \nonumber \\
k_B \Theta_{2} & = & \left[\frac{5}{3} \left\langle ( \hbar \omega)^2 
                     \right\rangle \right]^{1/2},
\end{eqnarray}
where $\langle \rangle$ denotes an average over the normal mode
spectrum, also vary little from valley to valley.  (These averages
always exclude the three zero-frequency modes that correspond to
center of mass motion.)  Their values fall in the ranges
\begin{eqnarray}
\Theta_{-2} & = & 114 \pm 4  \ {\rm K} \nonumber \\
\Theta_{0} & = & 98.7 \pm 0.1  \ {\rm K} \nonumber \\
\Theta_{2} & = & 154.0 \pm 0.1 \ {\rm K.}
\label{sodtemps}
\end{eqnarray}
The larger uncertainty in $\Theta_{-2}$ arises because $\Theta_{-2}$
is very sensitive to the lowest part of the frequency distribution, and
thus to a small system size.

\begin{figure}
\includegraphics[100,50][300,400]{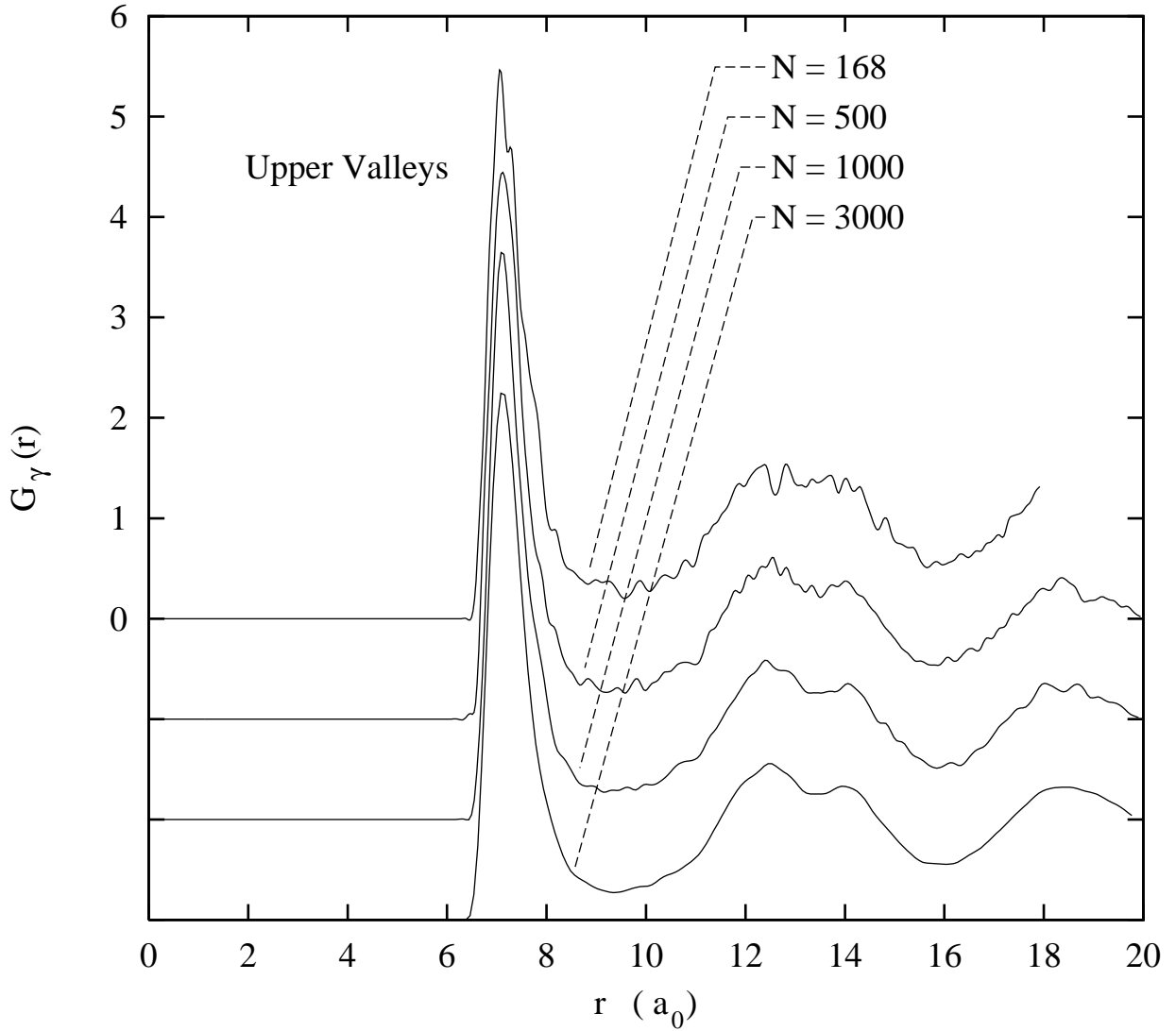}
\caption{Structure pair distribution functions $G_{\gamma}(r)$ for four 
         different valleys occupied by states from the upper group.  As 
         $N$ increases the small wiggles vanish.  Adapted from \cite{wall5}.}
\label{grrand}
\end{figure}

(2) The equilibrium configuration of particles at the bottom of a valley
is called a {\em structure}; Clements and Wallace denote the pair
distribution function for a structure $G_{\gamma}(r)$, where $\gamma$
labels the valley in which the structure lies.  They cooled several
confined states in the upper group to find the corresponding
structures and found that they all had very nearly the same
$G_{\gamma}(r)$, as illustrated in Figure \ref{grrand}.  The
fluctuations at small $N$ gradually vanish as $N$ increases.  (This
figure includes one valley at $N = 168$ which was studied in early
exploratory calculations, but which was not used in the final work
except at this point.)

(3) The universal $G_{\gamma}(r)$ for the valleys occupied by confined
upper states exhibits a split second peak (as is seen most clearly in
the $N = 3000$ plot in Figure \ref{grrand}), just as $g(r)$ for the
states from the lower group do [see point (6) above], but with the
first subpeak {\em higher} than the second.  Experiments on Ni, Co,
Cr, Fe, and Mn have identified this as a signature of an amorphous
structure \cite{davies, ichi, bennett, leung}.

(4) Clements and Wallace also constructed the set of Voronoi polyhedra
for each structure, and from this they computed the statistical
distributions of two coordination numbers: The number of faces per
polyhedron, and the angle between lines joining a particle to its
Voronoi neighbors.  They found that these distributions were universal
across all of the structures found by cooling states in the upper
group.

(5) The valleys occupied by states in the lower group, on the other
hand, do not exhibit universality in any of the properties measured:
Their depths, normal mode distributions, structure pair distribution
functions, and distributions of coordination numbers vary
substantially from valley to valley.

(6) The peaks in $G_{\gamma}(r)$ for any valley occupied by a state
from the lower group are more numerous and narrow than the peaks in
the universal $G_{\gamma}(r)$ of the valleys occupied by the upper
states, while the peaks in the bcc crystal $G_{\rm bcc}(r)$ are more
narrow still.

These results provide strong evidence that the many-body potential
surface of sodium contains two distinct classes of valleys: the
valleys in the first class exhibit universality in a wide variety of
properties, while the valleys in the second class don't.  The
potential surface is dominated by valleys of the first class, and
equilibrium states from the upper group are either confined to a
single valley of this class or move primarily among such valleys.
(That the first class dominates is shown by the fact that the system
almost always equilibrates to an upper state first.)  Since the upper
states lie along the same energy curve as the liquid states and $g(r)$
for the upper states evolves continuously into $g(r)$ for the liquid
as $T$ increases, the statistical mechanics of the liquid is
determined primarily by the properties of the first class of valleys.
We have less conclusive evidence that these valleys are approximately
harmonic [points (1) and (6) about the states] and that the structures
in the second class of valleys are less rigidly ordered than the bcc
crystal structure but more rigidly ordered than the structures in the
first class of valleys [points (3) and (6) about the valleys].
Thus we conclude that the valleys in the first class are random, and
those in the second class are symmetric.  (The rms vibrational
frequency and period for a ``typical'' many-body valley in sodium in
Subsection \ref{transits} were computed for the random valleys.)
Remnant symmetry
in the lower valleys is also consistent with their lower $\Phi_0$.  In
this system, the $\Phi_0$ values of the symmetric valleys range from
the value for the bcc valley up to the universal value for the random
valleys; symmetric valleys could conceivably have even higher $\Phi_0$
values, but none were found in this study.  Thus we conclude that our
general picture of the potential surface of a monatomic liquid is
rather well confirmed for this element.

\subsubsection{Lennard-Jones argon}

We conducted a less exhaustive study of Lennard-Jones argon
\cite{wall6} in which we reproduced some of the results described
above for sodium.  One difficulty we found with LJ argon is an
interesting instability: The system has a threshold density which lies
between the experimental densities of liquid Ar and fcc crystal Ar at
1 bar, and if one attempts to cool the system at a constant density
below this threshold, the system will collapse spontaneously until the
threshold is reached.  This limits one's ability to study densities
below the threshold at low temperatures.  Nonetheless, we were able to
explore many valleys at the threshold density using the same technique
as in sodium, and we found a group of states lying above those
occupying the fcc valley on a $\langle \Phi / N \rangle$ versus
$\langle \mbox{$\cal{K}$} / N \rangle$ graph, as in point (1) for
sodium.  The valleys occupied by these states exhibit the following
properties:

(1) They pass both tests of harmonicity (points (1) and (6) for states 
of sodium).

(2) Equilibrium states that move among these valleys are continuous with
the liquid states on the $\langle \Phi / N \rangle$ versus 
$\langle \mbox{$\cal{K}$} / N \rangle$ graph.

(3) The values of $\Phi_0 / N$ for all valleys lie in the same narrow range. 

Given our experience with sodium, we conclude tentatively that we have
found the random valleys in Lennard-Jones argon.  Thus the rms
vibrational frequency and period for a ``typical'' many-body valley in
argon in Subsection \ref{transits} were computed from these valleys.
Further tests on argon and other liquids will be considered in Section
\ref{outlook}.

\section{Equilibrium statistical mechanics}
\label{eq}

Now let us use the picture to develop a first order approximation to
the statistical mechanics of a monatomic liquid.  The specific heat
data suggest that the departures from harmonicity in the liquid's
Hamiltonian may be treated as higher-order perturbations, at least for
purposes of equilibrium statistical mechanics, and we shall keep this
in mind as we investigate the Hamiltonian and compute thermodynamic
quantities.  Corrections beyond the leading order will be considered
as we proceed.

\subsection{The Hamiltonian}
\label{Ham}

The general Hamiltonian for the system is written
\begin{equation}
H = \sum_{Ki} \frac{p_{Ki}^2}{2M} + \Phi(\{\mbox{\boldmath $r$}_K\})
\label{origH}
\end{equation}
where the index $K$ labels the particles, $i$ labels the components of
the position or momentum of a single particle, $M$ is the mass of one
atom, and $\Phi$ is the many-body potential.  We have argued that the
potential surface is dominated by a collection of nearly harmonic
valleys; let these valleys be labeled with the index $\gamma$, which
presumably runs from $1$ to approximately $w^N$.  We wish to consider
the form of the Hamiltonian when the system is localized in a
particular valley.  The coordinates of the particles at the valley
bottom will be denoted $\{\mbox{\boldmath $R$}_K(\gamma)\}$, and we
define
\begin{equation}
\mbox{\boldmath $u$}_K(\gamma) = \mbox{\boldmath $r$}_K - 
                                 \mbox{\boldmath $R$}_K(\gamma)
\label{disp}
\end{equation}
to be the displacement of the $K$th particle from its equilibrium
position.  The many-body potential in the valley will be denoted
$\Phi_{\gamma}$ and can be expanded as
\begin{equation}
\Phi_{\gamma}(\{\mbox{\boldmath $u$}_K(\gamma)\}) = \Phi_0(\gamma) + 
\frac{1}{2} \sum_{Ki,Lj} \Phi_{Ki,Lj}(\gamma)\ u_{Ki}(\gamma)\ u_{Lj}(\gamma)
+ \Phi_A(\gamma) 
\end{equation}
where 
\begin{equation}
\Phi_0(\gamma) = \Phi(\{\mbox{\boldmath $R$}_K\}) = 
                 \Phi(\{\mbox{\boldmath $u$}_K = \mbox{\boldmath $0$}\}),
\end{equation}
\begin{equation}
\Phi_{Ki,Lj}(\gamma) = \frac{\partial^2 \Phi}{\partial r_{Ki}\, \partial 
                       r_{Lj}}(\{\mbox{\boldmath $R$}_K\}),
\end{equation}
and $\Phi_A(\gamma)$ contains all of the higher order contributions to
$\Phi_{\gamma}$.  $\Phi_{Ki,Lj}(\gamma)$ is called the ``dynamical
matrix'' of the potential valley.  The Hamiltonian in the valley will
be denoted $H_{\gamma}$ and can now be written
\begin{equation}
H_{\gamma} = \Phi_0(\gamma) + H_H(\gamma) + \Phi_A(\gamma)
\label{fullH}
\end{equation}
where
\begin{equation}
H_H(\gamma) = \sum_{Ki} \frac{p_{Ki}^2}{2M} + \frac{1}{2} \sum_{Ki,Lj} 
              \Phi_{Ki,Lj}(\gamma)\ u_{Ki}(\gamma)\ u_{Lj}(\gamma)
\end{equation}
is the harmonic contribution.  An appropriate orthogonal transformation 
replaces the $\mbox{\boldmath $u$}_K(\gamma)$ with new coordinates 
$q_{\lambda}(\gamma)$ that diagonalize the dynamical matrix:
\begin{equation}
H_H(\gamma) = \sum_{\lambda} \left( \frac{p_{\lambda}^2}{2M} + \frac{1}{2}  
              M \omega_{\lambda}^2(\gamma)\,q_{\lambda}^2(\gamma) \right).
\label{harmH}
\end{equation}
This also defines the normal mode frequencies
$\omega_{\lambda}(\gamma)$.  If the system contains $N$ particles,
then $\lambda$ ranges from $1$ to $3N$.  If the valley happens to be
random, the Hamiltonian simplifies further; since the random valleys
all have the same depth and normal mode spectrum, the label $\gamma$
on the frequencies and $\Phi_0$ can be dropped, so
\begin{equation}
H_{\gamma} = \Phi_0 + H_H(\gamma) + \Phi_A(\gamma)
\label{fullHran}
\end{equation}
where
\begin{equation}
H_H(\gamma) = \sum_{\lambda} \left( \frac{p_{\lambda}^2}{2M} + \frac{1}{2}  
              M \omega_{\lambda}^2\,q_{\lambda}^2(\gamma) \right).
\label{harmHran}
\end{equation}
The Hamiltonian in Equation (\ref{fullH}), with the harmonic part
given by Equation (\ref{harmH}), is the starting point of our
treatment of equilibrium statistical mechanics.  Note that these
equations describe a restriction of the full Hamiltonian, Equation
(\ref{origH}), to a single potential valley, so they are defined only
within that valley.  The term $\Phi_A$ in the potential may describe
any sort of anharmonicity within the valley, but its main contribution
is expected to occur at the edges of the valley, where the potential
presumably flattens out (and departs from strict harmonic behavior)
before dipping down into a neighboring valley.

\subsection{The partition function}

We will now compute the quantum mechanical partition function and the
resulting thermodynamics, excluding exchange effects.  (A quantum
treatment is necessary for light elements including Li, Ne, and Be,
but without exchange effects this treatment will be insufficient for
describing liquid He.)  We will also display the classical limits of
our results; a fully classical development may be found in \cite{wall1}.

The partition function may be written
\begin{equation} 
Z = {\rm Tr}(e^{-\beta H}) = \sum_E g(E)\,e^{-\beta E}
\end{equation}
where $E$ ranges over the eigenvalues of the Hamiltonian $H$ and
$g(E)$ is a degeneracy factor which equals the dimension of the
eigenspace corresponding to $E$.  If the Hamiltonian described a
single harmonic valley of unbounded spatial extent with normal mode
frequencies $\omega_{\lambda}$, the eigenvalues would take the form
\begin{equation}
E = \sum_{\lambda} \left( n_{\lambda} + \frac{1}{2} \right) \hbar 
    \omega_{\lambda}
\label{eig}
\end{equation}
where the $n_{\lambda}$ are arbitrary nonnegative integers.  We have
argued that the true potential is dominated overwhelmingly by a single
class of nearly harmonic valleys, the random valleys, all of which
have the same normal mode spectrum; therefore, let us approximate the
eigenvalues of the harmonic part of the actual Hamiltonian, Equation
(\ref{harmH}), with values of the form given in Equation (\ref{eig})
using the universal random valley normal mode spectrum; it then
remains to determine the degeneracy of each.  Our approximation
suggests the existence of eigenfunctions of $H$ which are largely
confined to individual valleys; clearly there would be approximately
$w^N$ of these for each eigenvalue, one per valley, and these would be
approximately orthogonal (because they are almost spatially disjoint),
hence approximately linearly independent.  Thus we suggest $g(E)
\approx w^N$ independent of $E$, or
\begin{eqnarray}
Z & \approx & \sum_{\{n_{\lambda}\}} w^N \exp\left(-\beta \left[ \Phi_0 + 
              \sum_{\lambda} \left( n_{\lambda} + \frac{1}{2} \right) \hbar 
              \omega_{\lambda} \right] \right) \nonumber \\
 & = & w^N e^{-\beta \Phi_0} \sum_{\{n_{\lambda}\}} \prod_{\lambda} 
       \exp\left[-\beta \left( n_{\lambda} + \frac{1}{2} \right) \hbar 
       \omega_{\lambda} \right] \nonumber \\
 & = & w^N e^{-\beta \Phi_0} \prod_{\lambda} \sum_n \exp\left[-\beta 
       \left( n + \frac{1}{2} \right) \hbar \omega_{\lambda} \right] 
       \nonumber \\
 & = & w^N e^{-\beta \Phi_0} \prod_{\lambda} \frac{e^{-\frac{1}{2}\beta
       \hbar\omega_{\lambda}}}{1 - e^{-\beta\hbar\omega_{\lambda}}}.
\end{eqnarray}
This is the approximate partition function for the liquid.  In the
classical limit ($\hbar \omega_{\lambda} \ll k_B T$ for all $\lambda$),
\begin{equation}
Z \approx w^N e^{-\beta \Phi_0} \prod_{\lambda} (\beta \hbar 
          \omega_{\lambda})^{-1}.
\end{equation}
We have made three noteworthy approximations in calculating $Z$:
First, we have neglected the contributions from the symmetric and
crystalline valleys.  This is a superb approximation, however, since
the random valleys vastly outnumber the other two types.  Second, we
have neglected the term $\Phi_A$ in the Hamiltonian (\ref{fullH}).
Third, in using energy eigenvalues of the form (\ref{eig}) we have
implicitly extended a single potential valley throughout all of
configuration space, failing to take into account its limited spatial
extent; we have thus neglected the existence of {\em boundaries} of
the valleys.  These last two approximations are more significant, and
their effects will be included in our subsequent calculations.

\subsection{Thermodynamic state functions}
\label{thermo}

To each thermodynamic state function $X$ we will append a term of the
form $X_{AB}$ representing the corrections due to anharmonicity and
boundary effects, as discussed immediately above, without further
comment.  The Helmholtz free energy is
\begin{eqnarray}
F & = & -k_B T \ln Z \nonumber \\
  & = & \Phi_0 - N k_B T \ln w + \sum_{\lambda} \left[\frac{1}{2} \hbar 
        \omega_{\lambda} - k_B T \ln(n_{\lambda} + 1)\right] + F_{AB} 
\label{F}
\end{eqnarray}
where 
\begin{equation}
n_{\lambda} = \frac{1}{e^{\beta\hbar\omega_{\lambda}} - 1},
\end{equation}
the entropy is
\begin{eqnarray}
S & = & -\left(\frac{\partial F}{\partial T}\right)_V \nonumber \\
  & = & N k_B \ln w + k_B \sum_{\lambda} \left[ (n_{\lambda} + 1) 
        \ln (n_{\lambda} + 1) - n_{\lambda} \ln n_{\lambda} \right] + S_{AB} 
\end{eqnarray}
where $n_{\lambda}$ is defined as before, and the internal energy is
\begin{eqnarray}
U & = & F + TS \nonumber \\
  & = & \Phi_0 + \sum_{\lambda} \left(n_{\lambda} + \frac{1}{2}\right)
        \hbar\omega_{\lambda} + U_{AB}.
\end{eqnarray}
Finally, the constant-volume specific heat is
\begin{eqnarray}
C_V & = & \left(\frac{\partial U}{\partial T}\right)_V \nonumber \\
    & = & k_B \sum_{\lambda} \left[ n_{\lambda} (n_{\lambda} + 1) 
          (\beta\hbar\omega_{\lambda})^2 \right] + C_{AB}.
\end{eqnarray}

It is convenient to express the state functions in the classical limit
in terms of the temperature $\Theta_0$ defined as in Subsection
\ref{structure} by
\begin{equation}
\ln k_B \Theta_0 = \frac{\sum_{\lambda} \ln \hbar \omega_{\lambda}}{3N}.
\end{equation} 
Using this definition, in the limit $\hbar \omega_{\lambda} \ll k_B T$
for all $\lambda$ we find
\begin{eqnarray}
F & = & \Phi_0 - Nk_BT\ln w - 3Nk_BT\ln(T/\Theta_0) + F_{AB}, \label{classF} \\
S & = & Nk_B\ln w + 3Nk_B[\ln(T/\Theta_0) + 1] + S_{AB}, \label{classS} \\
U & = & \Phi_0 + 3Nk_BT + U_{AB}, \label{classU} \\
C_V & = & 3Nk_B + C_{AB}. \label{classC}
\end{eqnarray}

\subsection{Comparison with experiment}
\label{exp}

All comparisons with experimental data will be done in the classical
limit.  First, we derive the expression for entropy of melting.
The entropy for a monatomic harmonic crystal has the same form as the
liquid, Equation (\ref{classS}), without the $Nk_B \ln w$ term
since the system resides in a single potential valley.  Let the
superscript $l$ denote quantities of the liquid and $c$ those of the
crystal; then
\begin{eqnarray}
S^l & = & Nk_B\ln w + 3Nk_B[\ln(T/\Theta_0^l) + 1] + S_{AB}^l + S_E^l 
          \nonumber \\
S^c & = & 3Nk_B[\ln(T/\Theta_0^c) + 1] + S_A^c + S_E^c 
\label{crysliqS}
\end{eqnarray}
where $S_E$ is the valence electronic contribution to the entropy
(note that the crystal's entropy has anharmonic corrections but no
boundary corrections), so the entropy of melting at constant density
$\Delta S$ is given by
\begin{eqnarray}
\Delta S & = & S^l(T_m) - S^c(T_m) \nonumber \\
         & = & Nk_B\ln w + 3Nk_B\ln(\Theta_0^c/\Theta_0^l) + \left(S_{AB}^l -
               S_{A}^c\right) + \left(S_E^l - S_E^c\right).
\label{DeltaS}
\end{eqnarray}
Let us consider a normal melting element.  Since its electronic
structure is not changed significantly upon melting, it is reasonable
to suspect that $S_E^l \approx S_E^c$, so assuming anharmonic and
boundary effects are small, the entropy of melting is dominated by the
first two terms, the second of which depends strongly on the
individual element and the first of which may or may not depend
strongly, depending on how $w$ varies between different substances.
The experimental data from Subsection \ref{entmelt} reveal that
$\Delta S = 0.8\,Nk_B$ for all nearly-free-electron metals with a
small scatter; the only term in Equation (\ref{DeltaS}) that could
reasonably be considered universal and thus account for these data is
the first, assuming $w$ is itself universal and $\ln w = 0.8$.  That
in turn implies that $\Theta_0^c \approx \Theta_0^l$ for normal
melters, with the departures from strict equality, along with
anharmonic, boundary, and electronic entropy contributions, accounting
for the scatter in $\Delta S$.  We have verified the prediction
$\Theta_0^c \approx \Theta_0^l$ for sodium and Lennard-Jones argon,
both of which are normal melters; for sodium in the bcc crystal phase
\cite{wall4}
\begin{equation} \Theta_0 = 99.65 \ {\rm K}, \end{equation}
which is quite close to $\Theta_{0} = 98.7 \pm 0.1$ K for the liquid
[Equation (\ref{sodtemps})].  For argon, $\Theta_0 = 42.5$ K for the
liquid at $\rho = 1.6000$ g/cm$^3$, and $\Theta_0 = 43.4$ K for the
fcc crystal at the same density \cite{wall6}.  We also predict that
the much higher $\Delta S$ values of the anomalous melters can be
accounted for mainly from the different values of $\Theta_0$ and $S_E$
for the two phases, both because of their very different electronic
structures, plus the usual small anharmonic and boundary effects.

\begin{figure}
\includegraphics*[120,195][700,700]{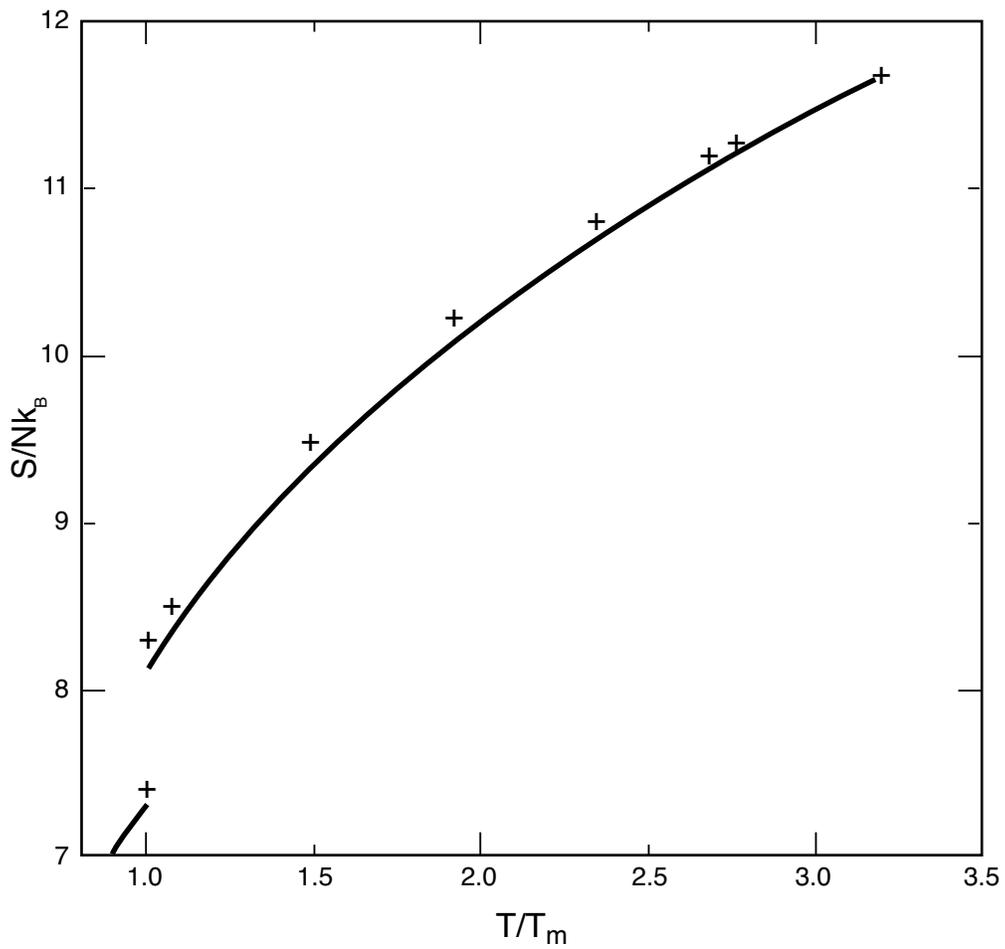}
\caption{Theoretical prediction of the entropy of mercury in crystal 
         and liquid phases (curve) compared with experimental data 
         (crosses).  Adapted from \cite{wall1}.}
\label{SHg}
\end{figure}

\begin{figure}
\includegraphics*[100,195][700,700]{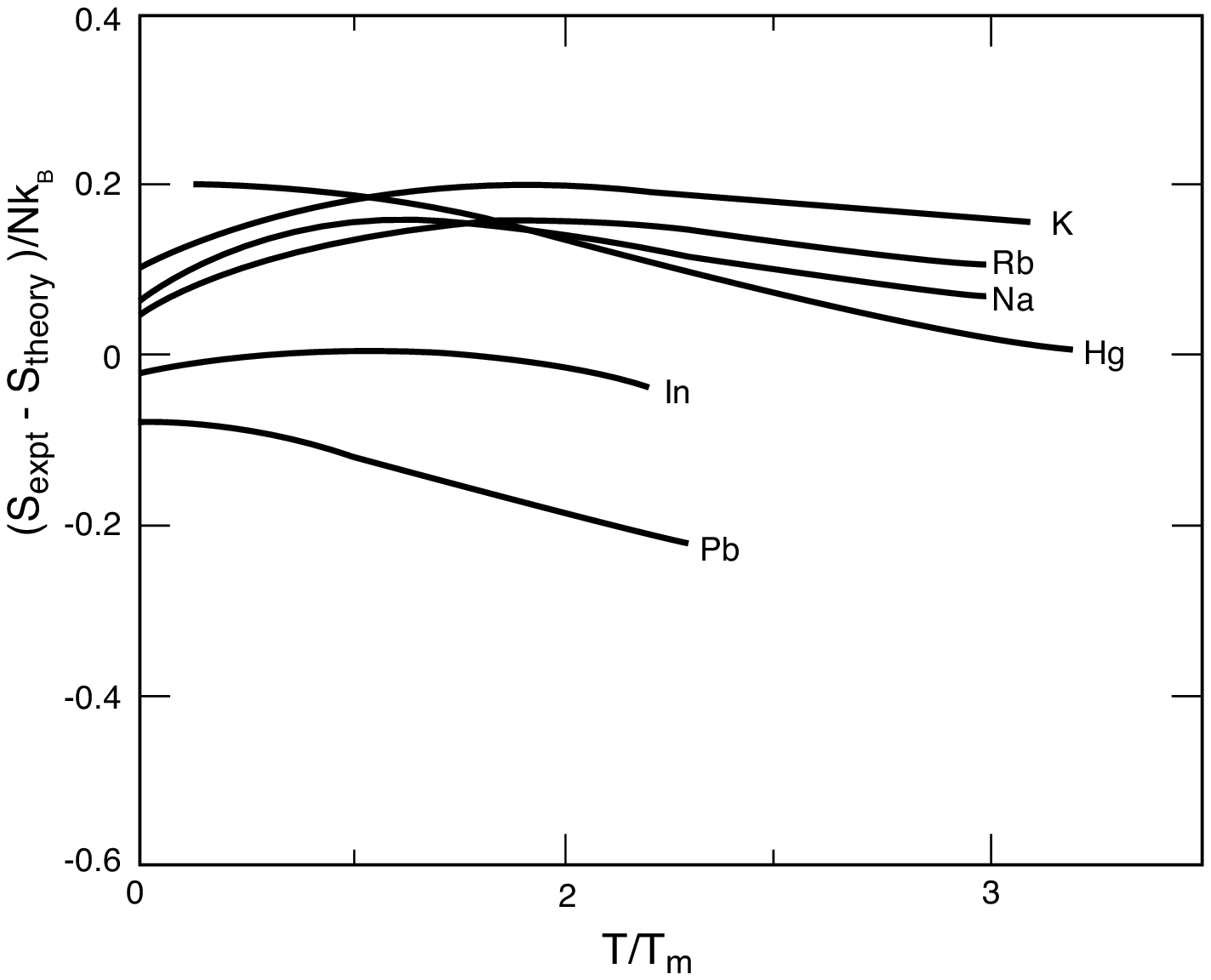}
\caption{Difference between experimental and theoretical entropy for six 
         liquid metals.  Adapted from \cite{wall1}.}
\label{Sdiff}
\end{figure}

Second, Wallace \cite{wall1} has compared Equations (\ref{crysliqS})
(neglecting anharmonic and boundary terms) to experimental entropy
data for six nearly-free-electron metals; the criteria used to
select the six elements, and the details of the correction of the data
for density changes, are given in \cite{wall1}.  Figure \ref{SHg}
shows the theoretical prediction for the entropy of mercury in crystal
and liquid phases, over a temperature range from below $T_m$ to
$3.2\,T_m$, compared to experimental data.  The agreement is most
encouraging.
The differences between experimental and theoretical entropy as a
function of $T/T_m$ for all six elements are shown in Figure
\ref{Sdiff}.  The differences fall within the expected errors in the
analysis, as discussed in \cite{wall1}.
Recall that all thermodynamic functions are derived from a single
potential, taken in this case to be the free energy, which has both a
zero-temperature part, $\Phi_0$, and a thermal part (see Equation
(\ref{F}) or (\ref{classF})); because of this, it suffices to compare one 
function (in this case, entropy) with experiment to guarantee the 
accuracy of the thermal part of our entire thermodynamic treatment.

To check the zero-temperature part of our thermodynamics, we consider
one further point of contact with experiment.  As Wallace has shown in
\cite{wall1}, the potential minima of the liquid and crystal,
corrected for density differences, should obey the relation
\begin{equation} 
\Phi_0^l(\rho_{lm}) - \Phi_0^c(\rho_{lm}) \approx T_m \Delta S
\end{equation}
where $\rho_{lm}$ is the density of the liquid at melt.  This assumes
that both liquid and crystal obey the harmonic relation $\langle \Phi
\rangle = \Phi_0 + (3/2)Nk_BT$, so the difference in $U = \langle \Phi
\rangle + (3/2)Nk_BT$ equals the difference in $\Phi_0$.  However, as
noted in \cite{wall4}, experiments on sodium have determined that $T_m
\Delta S$ = 1.7 mRy/atom, while MD calculations (Figure \ref{pvsk})
show that the difference in potential minima between the random
valleys and the bcc valley is roughly 0.92 mRy/atom, which is of the
right order of magnitude but is 46\% smaller than experiment.  The
reason for this discrepancy is easy to find; as can be seen in Figure
\ref{pvsk} and more clearly in Figure 4 of \cite{wall4}, which plots
the states in the random valleys together with the liquid states, the
slope of the line followed by the states increases at higher $T$, so
$\langle \Phi \rangle$ does not obey the harmonic relation.  If one
uses the values of $\langle \Phi \rangle$ from Figure 4 of
\cite{wall4} at melt, one finds that the difference in $U$ is in fact
1.7 mRy/atom.  We will return to this anharmonicity in Section
\ref{outlook}.

\section{Nonequilibrium statistical mechanics}
\label{noneq}

It is not often enough emphasized that equilibrium statistical
mechanics and its nonequilibrium counterpart ultimately have the same
starting point, the Hamiltonian of the system, although they make use
of the Hamiltonian in very different ways.  Both begin with a
decomposition of the Hamiltonian into ``free'' and ``transition''
terms, where the ``free'' term is the more tractable of the two and
can be fairly readily diagonalized.  In equilibrium statistical
mechanics, the entire Hamiltonian contributes to the partition
function, but often the contribution of the transition term cannot be
computed exactly, so its effects are usually included as a
perturbative approximation.  Nonequilibrium statistical mechanics, on
the other hand, treats the system as executing transitions between the
states that diagonalize the free part of the Hamiltonian, with the
transition term determining cross sections and transition rates.  In a
gas, for example, the free part of the Hamiltonian describes $N$-body
free motion, while the transition part is responsible for
interparticle interactions.  An equilibrium treatment of a gas
portrays it as an ideal gas with perturbations away from ideal
behavior produced by the interaction terms; a nonequilibrium treatment
via the Boltzmann equation portrays the gas as executing transitions
between many-body free motion states by means of collisions, which are
ultimately mediated by the same interaction terms responsible for
perturbations in the equilibrium treatment.  The liquid is analogous:
The Hamiltonian decomposes into a ``free'' term, which is $\Phi_0$
plus the harmonic term from Equation (\ref{harmH}), and a
``transition'' term, which consists of $\Phi_A$ from Equation
(\ref{fullH}) plus the presence of boundaries.  The equilibrium
statistical mechanics of the liquid, as we have seen, is dominated by
the ``free'' term, although the other terms introduce corrections; and
nonequilibrium statistical mechanics ultimately should treat the
liquid as executing transits between states confined to individual
valleys, with transits mediated by the boundary term of the
Hamiltonian.  (This fact connects transits in nonequilibrium mechanics
to the boundary corrections in equilibrium mechanics; see Section
\ref{outlook}.)  Thus transits, which do not appear in the equilibrium
results at all to lowest order, will play a central role in
understanding the liquid's nonequilibrium behavior.  Because of this
we begin by writing the position of the $K\!$th particle in the liquid
as
\begin{equation}
\mbox{\boldmath $r$}_K(t) = \mbox{\boldmath $R$}_K(t) + 
                            \mbox{\boldmath $u$}_K(t)
\label{rdecomp}
\end{equation}
where $\mbox{\boldmath $R$}_K(t)$ is the location of the center about
which the particle oscillates between transits and $\mbox{\boldmath
$u$}_K(t)$ is motion about that center.  Then $\mbox{\boldmath
$R$}_K(t)$ changes only when a transit involving particle $K$ takes
place.  (Compare \mbox{Equation} (\ref{disp}).)  This decomposition, 
which is motivated by a corresponding decomposition of the Hamiltonian, 
is the starting point of our nonequilibrium treatment.

We will work in the linear regime, in which the coefficients
determining the system's nonequilibrium response (self-diffusion, bulk
viscosity, shear viscosity, thermal conductivity, etc.) are related to
equilibrium time-correlation functions by expressions of the Green-Kubo
form \cite{hansmac}; thus our goal is to understand the physics behind
the appropriate correlation functions.  We will perform a sample
calculation of two simple correlation functions, and then we will
proceed to the very important velocity autocorrelation function, which
determines the self-diffusion coefficient.  We will work in the
classical limit; quantum aspects will be discussed in Section
\ref{outlook}.

\subsection{Correlation functions in the absence of transits}

An important part of this work involves calculating correlation
functions of harmonically varying quantities, so we will first show
such a computation by considering the one-particle functions $\langle
\mbox{\boldmath $u$}(t) \cdot \mbox{\boldmath $u$}(0) \rangle$ and
$\langle \mbox{\boldmath $v$}(t) \cdot \mbox{\boldmath $v$}(0)
\rangle$ in the simplest situation, when the temperature is
sufficiently low that the system remains in a single valley without
transits.  (We recall from Subsection \ref{transits} that this is
below roughly $30$ K for sodium and $17.1$ K for Lennard-Jones argon.)
Ultimately we will be comparing these results with MD simulations, in
which the center of mass of the system is stationary; in this case,
only $N-1$ of the particles' positions are independent, so we define
correlation functions as averages over particles with that
restriction, and we divide by $N-1$, not $N$, to take into account the
reduced number of independent degrees of freedom.  We consider the
position correlation function first.
\begin{eqnarray}
\langle \mbox{\boldmath $u$}(t) \cdot \mbox{\boldmath $u$}(0) \rangle & 
        \equiv & \frac{1}{N-1} \sum_K \langle \mbox{\boldmath $u$}_K(t) \cdot 
        \mbox{\boldmath $u$}_K(0) \rangle \nonumber \\
 & = & \frac{1}{N-1} \sum_{Ki} \langle u_{Ki}(t)\,u_{Ki}(0) \rangle. 
\end{eqnarray}
Let the orthogonal transformation from the original coordinates to the
normal mode coordinates be denoted $w_{Ki,\lambda}$, so
\begin{equation}
u_{Ki}(t) = \sum_{\lambda} w_{Ki,\lambda}\,q_{\lambda}(t),
\end{equation}
where the normal modes are denoted $q_{\lambda}$ as in Subsection
\ref{Ham}; then 
\begin{eqnarray}
\langle \mbox{\boldmath $u$}(t) \cdot \mbox{\boldmath $u$}(0) \rangle & = &
 \frac{1}{N-1} \sum_{Ki,\lambda,\lambda'} \langle w_{Ki,\lambda}\,
 w_{Ki,\lambda'}\,q_{\lambda}(t)\,q_{\lambda'}(0) \rangle \nonumber \\
 & = & \frac{1}{N-1} \sum_{\lambda} \langle q_{\lambda}(t)\,q_{\lambda}(0) 
       \rangle.
\end{eqnarray}
Because the potential of the system is invariant under translations,
three of the $q_{\lambda}$ (the components of the center of mass) are
zero-frequency modes; since these modes are not excited by assumption,
$\lambda$ ranges from 1 to $3N-3$ over the nonzero modes.  Now
\begin{equation}
\langle q_{\lambda}(t)\,q_{\lambda}(0) \rangle \equiv 
\left\langle \left\{e^{-i\mbox{$\cal{L}$}t}q_{\lambda}\right\}q_{\lambda} 
\right\rangle
\end{equation}
where $\cal{L}$ is the Liouville operator for the system, so in our 
harmonic approximation
\begin{eqnarray}
\langle q_{\lambda}(t)\,q_{\lambda}(0) \rangle & = & \left\langle 
        \left\{q_{\lambda} \cos(\omega_{\lambda}t) + \frac{p_{\lambda}}
        {M\omega_{\lambda}} \sin(\omega_{\lambda}t) \right\}q_{\lambda} 
        \right\rangle \nonumber \\
 & = & \langle q_{\lambda}^2 \rangle \cos(\omega_{\lambda}t) + 
       \frac{\langle q_{\lambda}\,p_{\lambda} \rangle}{M\omega_{\lambda}} 
       \sin(\omega_{\lambda}t)
\end{eqnarray}
where $\{\omega_{\lambda}\}$ is the set of normal mode frequencies of
a random valley.  (The system is overwhelmingly likely to be in a
random valley because such valleys dominate the potential surface.)
The two averages are easily calculated in the canonical ensemble,
\begin{equation}
\langle q_{\lambda}^2 \rangle = \frac{k_B T}{M \omega_{\lambda}^2}, \ \ \ 
\langle q_{\lambda}\,p_{\lambda} \rangle = 0, 
\label{avg}
\end{equation}
and the final result is
\begin{equation}
\langle \mbox{\boldmath $u$}(t) \cdot \mbox{\boldmath $u$}(0) \rangle =
\frac{1}{N-1} \frac{k_B T}{M} \sum_{\lambda} \frac{\cos(\omega_{\lambda}t)}
{\omega_{\lambda}^2}.
\label{uu}
\end{equation}
(Note that, aside from the fact that $\mbox{\boldmath $u$}(0)$ and
$\mbox{\boldmath $u$}(t)$ would appear symmetrically in the definition
of the correlation function, the quantum calculation proceeds
identically to its classical counterpart until Equation (\ref{avg}).)
Since $\mbox{\boldmath $v$}(t) = \dot{\mbox{\boldmath $u$}}(t)$ in the
absence of transits, a similar line of reasoning leads to
\begin{equation}
\langle \mbox{\boldmath $v$}(t) \cdot \mbox{\boldmath $v$}(0) \rangle =
\frac{1}{N-1} \frac{k_B T}{M} \sum_{\lambda} \cos(\omega_{\lambda}t).
\label{vv}
\end{equation} 
These results, which have the same form as the corresponding results
for a harmonic crystal, will serve as a reference point for the work
of the next subsection.

\subsection{Velocity autocorrelation function and diffusion coefficient}

Now we will consider the velocity autocorrelation function $Z(t)$, defined 
by
\begin{equation} 
Z(t) = \frac{1}{3} \langle \mbox{\boldmath $v$}(t) \cdot 
       \mbox{\boldmath $v$}(0) \rangle,
\end{equation}
which determines the self-diffusion coefficient $D$ through the
Green-Kubo relation \cite{hansmac}
\begin{equation} D = \int_0^\infty Z(t)\,dt. \end{equation}
We predict from Equation (\ref{vv}) that at sufficiently low temperatures
\begin{equation}
Z(t) = \frac{1}{3N-3} \frac{k_B T}{M} \sum_{\lambda} \cos(\omega_{\lambda}t),
\label{lowTZ}
\end{equation}
or in terms of $\hat{Z}(t) \equiv Z(t) / Z(0)$,
\begin{equation}
\hat{Z}(t) = \frac{1}{3N-3} \sum_{\lambda} \cos(\omega_{\lambda}t).
\label{lowTZhat}
\end{equation}
It is not at all clear, however, how this result will be modified at
higher temperatures by transits; certainly even an approximate
solution to the system's equations of motion seems well out of reach.
In \cite{wall8} we argued that trying to understand the motion of the
system in terms of a normal mode decomposition would be unhelpful once
transits begin to occur for the following reasons: Over a broad range
of temperatures, we suspect that any given particle will participate
in a transit roughly once per mean vibrational period (this will be
verified below), so each particle will experience roughly ten transits
by its neighbors per period.  Each such transit changes the many-body
valley in which the system lies, thus changing the particular normal
mode decomposition in which the coordinates of all the particles are
expressed.  Perhaps such a change would minimally affect the
coordinates of far away particles, but it should certainly have a
substantial effect on the coordinates of the near neighbors.  In
response to this, one could instead suggest that the normal mode
picture needs only to be supplemented, not replaced, and this line of
reasoning has been followed most notably in some INM work (for
example, \cite{keyes1, keyes2}); however, that work has focused not on
constructing an explicit model for the system's motion while
transiting, but on modeling the effects of transits on Equation
(\ref{lowTZ}) directly in the general form suggested by Zwanzig
\cite{zwan}.  If we must resort to models, we strongly prefer
developing a model of the actual motion of the particles in the
liquid, transits included, and then calculating $Z(t)$ from there,
because we believe that the important thing to be understood is the
motion, not just the behavior of one correlation coefficient, and
because such a model can then be used to calculate any single-particle
correlation coefficient one chooses.  Thus in \cite{wall8} we proposed
a mean-atom-trajectory model, which consists of a single average
particle in the liquid periodically transiting between single-particle
equilibrium positions while executing harmonic motion between
transits.  We then incorporated into this model the essential features
one expects from an actual solution to the equations of motion of the
system, as shown below.

Since each transit carries the system with overwhelming likelihood
between random valleys, it is sensible to model the average particle's
motion between transits in terms of oscillations at the random valley
frequency distribution, or
\begin{eqnarray}
\mbox{\boldmath $r$}(t) & = & \mbox{\boldmath $R$} + \mbox{\boldmath $u$}(t)
                              \nonumber \\
& = & \mbox{\boldmath $R$} + \sum_{\lambda} \mbox{\boldmath $w$}_{\lambda}
                             \sin(\omega_{\lambda}t + \alpha_{\lambda})  
\label{rt}
\end{eqnarray}
where \mbox{\boldmath $R$} and \mbox{\boldmath $u$}$(t)$ are the
mean-atom equivalents of \mbox{\boldmath $R$}$_K$ and \mbox{\boldmath
$u$}$_K(t)$ from Equation (\ref{rdecomp}).  (Between transits
\mbox{\boldmath $R$} has no time dependence.)  Now the parameters
$\mbox{\boldmath $w$}_{\lambda}$ and $\alpha_{\lambda}$ in
$\mbox{\boldmath $u$}(t)$ remain to be determined.  Let us assume that
the values of the phases $\alpha_{\lambda}$ are randomly distributed
among the particles; then one calculates $Z(t)$ from Equation
(\ref{rt}) by differentiating to find $\mbox{\boldmath $v$}(t)$,
computing the product $\mbox{\boldmath $v$}(t) \cdot \mbox{\boldmath
$v$}(0)$, and averaging over each of the $\alpha_{\lambda}$
separately; the result is
\begin{equation}
Z(t) = \frac{1}{6} \sum_{\lambda} |\mbox{\boldmath $w$}_{\lambda}|^2 
       \omega_{\lambda}^2 \cos(\omega_{\lambda}t).
\label{kindaZ}
\end{equation}
Equation (\ref{kindaZ}) becomes Equation (\ref{lowTZ}) with the choice
\begin{equation}
\mbox{\boldmath $w$}_{\lambda} = \sqrt{\frac{1}{N-1}\frac{2k_B T}
                                 {M\omega_{\lambda}^2}}
                                 \,\hat{\mbox{\boldmath $w$}}_{\lambda}
\label{wchoice}
\end{equation}
where $\hat{\mbox{\boldmath $w$}}_{\lambda}$ is an arbitrarily chosen
unit vector.  Thus Equation (\ref{rt}) with the phases
$\alpha_{\lambda}$ randomly chosen and $\mbox{\boldmath
$w$}_{\lambda}$ given by Equation (\ref{wchoice}), with the unit
vectors $\hat{\mbox{\boldmath $w$}}_{\lambda}$ also randomly chosen,
constitute our mean-atom-trajectory model between transits.  (A brief
calculation shows that this model also yields the correct result for
$\langle \mbox{\boldmath $u$}(t) \cdot \mbox{\boldmath $u$}(0)
\rangle$ from Equation (\ref{uu}).)

Next we must determine the effect of transits on the parameters in
\mbox{\boldmath $r$}$(t)$, and that requires an explicit model of both
the transit of an average particle and the rate at which transits
occur.  First, the transit process itself.  We assume that the transit
occurs instantaneously (the particle simply crosses the surface
separating distinct valleys), so it must conserve both the particle's
position $\mbox{\boldmath $r$}(t)$ and velocity $\mbox{\boldmath
$v$}(t)$.  To be more specific, we assume that the transit occurs in
the forward direction, so that the center of the new valley lies an
equal distance away from the particle but on the opposite side from
the center of the old valley.  Let $\mbox{\boldmath $r$}^{\rm
before}(t)$, $\mbox{\boldmath $R$}^{\rm before}$, and $\mbox{\boldmath
$u$}^{\rm before}(t)$ be the position parameters from Equation
(\ref{rt}) before the transit, and let $\mbox{\boldmath $r$}^{\rm
after}(t)$, $\mbox{\boldmath $R$}^{\rm after}$, and $\mbox{\boldmath
$u$}^{\rm after}(t)$ be the parameters after; then our assumption of a
forward transit implies that $\mbox{\boldmath $u$}^{\rm after}(t) =
-\mbox{\boldmath $u$}^{\rm before}(t)$, and this together with
$\mbox{\boldmath $r$}^{\rm before}(t) = \mbox{\boldmath $r$}^{\rm
after}(t)$ implies
\begin{equation}
\mbox{\boldmath $R$}^{\rm after} = \mbox{\boldmath $R$}^{\rm before} 
                            + 2\mbox{\boldmath $u$}^{\rm before}(t).
\end{equation}
This is the change in \mbox{\boldmath $R$} produced by a transit.  We
choose to leave the unit vectors $\hat{\mbox{\boldmath
$w$}}_{\lambda}$ in Equation (\ref{wchoice}) unaffected by transits,
leaving only the effect on the phases $\alpha_{\lambda}$ to be
determined.  They must change in such a way as to reverse the sign of
$\mbox{\boldmath $u$}(t)$ but conserve $\mbox{\boldmath $v$}(t)$;
since $\mbox{\boldmath $u$}(t)$ is a sum of sines while
$\mbox{\boldmath $v$}(t)$ is a sum of cosines, this is easily done by
reversing the signs of the arguments $(\omega_{\lambda}t +
\alpha_{\lambda})$ in Equation (\ref{rt}).  Let the transit occur at time
$t_{0}$; then $\omega_{\lambda}t_{0} + \alpha_{\lambda}^{\rm after} =
-(\omega_{\lambda}t_{0} + \alpha_{\lambda}^{\rm before})$ so
\begin{equation}
\alpha_{\lambda}^{\rm after} = -2\omega_{\lambda}t_{0} -
                         \alpha_{\lambda}^{\rm before}.
\end{equation}
Thus, a transit is implemented at time $t_{0}$ by leaving the
$\hat{\mbox{\boldmath $w$}}_{\lambda}$ alone and making the substitutions
\begin{eqnarray}
\mbox{\boldmath $R$} & \rightarrow & \mbox{\boldmath $R$} + 2\mbox{\boldmath
                                                 $u$}(t_{0}) \nonumber \\
\alpha_{\lambda} & \rightarrow & -2\omega_{\lambda}t_{0} - \alpha_{\lambda}.
\label{imptrans}
\end{eqnarray}
This conserves {\boldmath $r$}$(t)$, reverses the sign of {\boldmath
$u$}$(t)$, and conserves {\boldmath $v$}$(t)$.

Let the temperature-dependent rate at which transits occur be denoted
$\nu$, so in small time interval $\Delta t$ a transit occurs with
probability $\nu \Delta t$.  As a transition rate, $\nu$ would ideally
be calculated from matrix elements of the term in the Hamiltonian
responsible for transits using Fermi's Golden Rule, and we will revisit
this possibility in Section \ref{outlook}, but for now we will take
the simpler path of fitting $\nu$ to the results of MD simulations.

Now the model consists of two parts.  (a) Between transits, the
average particle oscillates as given by Equations (\ref{rt}) and
(\ref{wchoice}), with the phases $\alpha_{\lambda}$ and unit vectors
$\hat{\mbox{\boldmath $w$}}_{\lambda}$ assigned randomly.  (b) In each
small time interval $\Delta t$ a transit occurs with probability $\nu
\Delta t$; if it occurs, it replaces \mbox{\boldmath $R$} and the
$\alpha_{\lambda}$ with new values according to Equation
(\ref{imptrans}).  With the addition of transits, we can no longer
express {\boldmath $r$}$(t)$ and {\boldmath $v$}$(t)$ in closed form
at all times, so we no longer have a closed form for $Z(t)$; but the
model can be implemented easily on a computer, and then the data from
the run can be used to calculate $Z(t)$ and $\hat{Z}(t)$ in a manner 
analogous to an MD simulation.  

\begin{figure}
\includegraphics{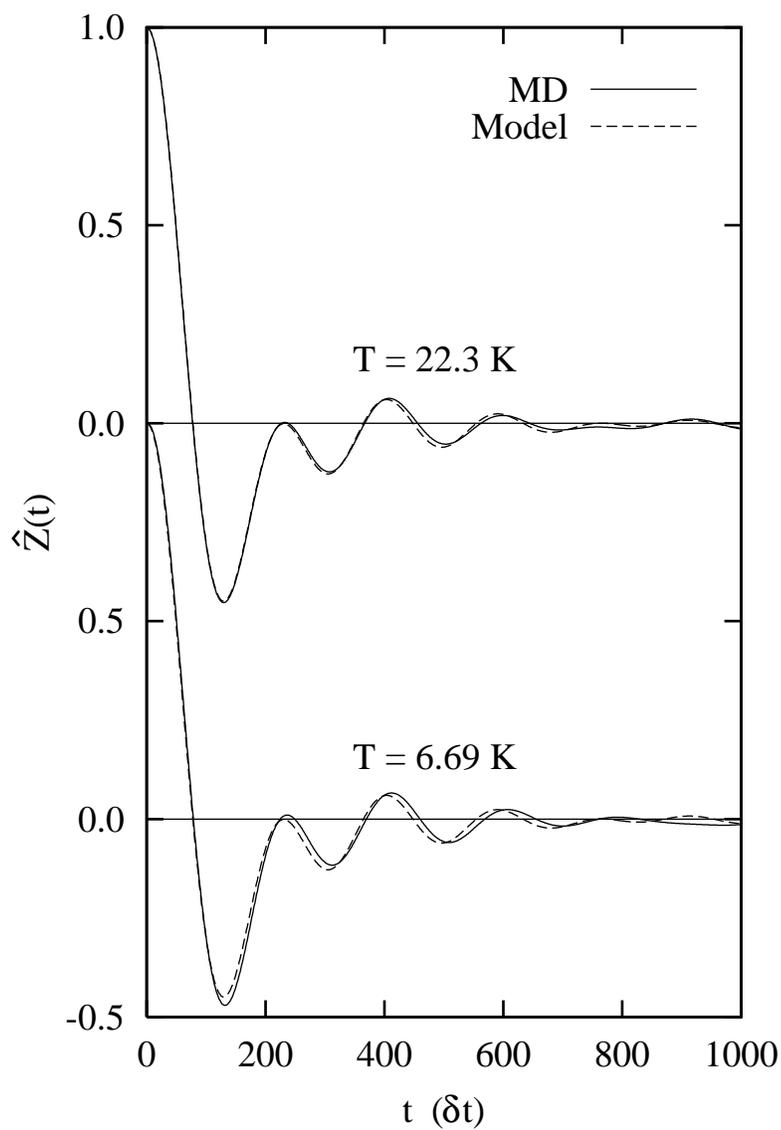}
\caption{The model prediction for $\hat{Z}(t)$ at $\nu = 0$ compared with 
         MD results for glassy liquid Na at $T = 6.69$ K and $T = 22.3$ K.  
         From \cite{wall8}.}
\label{MDvsmodel}
\end{figure}

\begin{figure}
\includegraphics{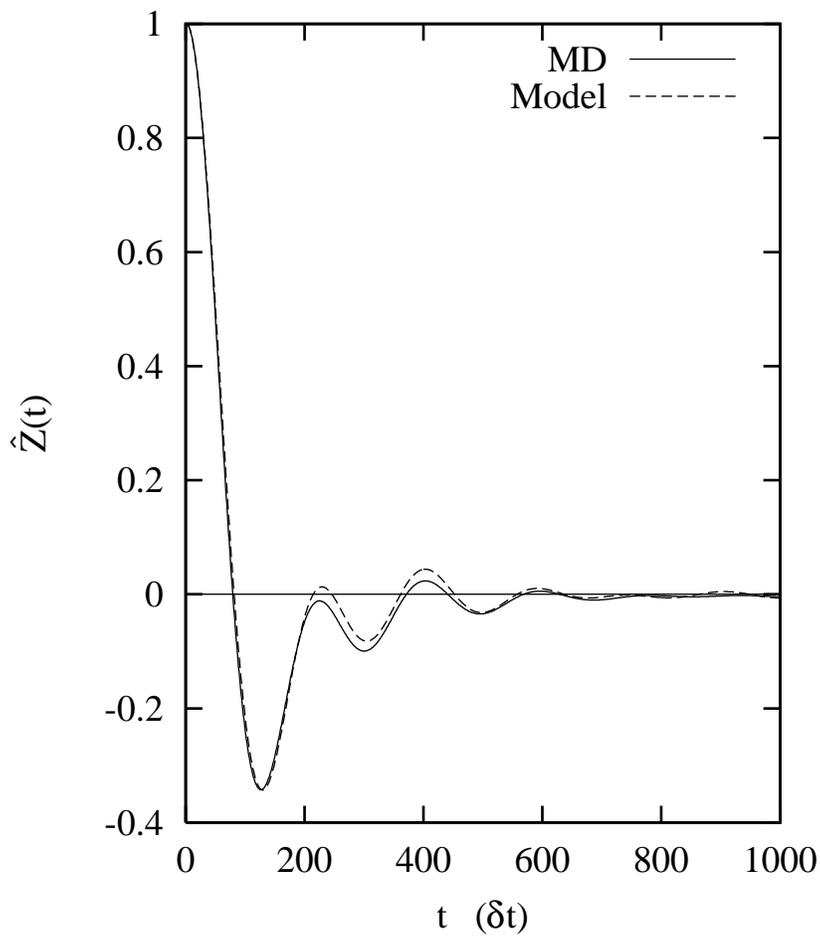}
\caption{The model prediction for $\hat{Z}(t)$ at $\nu = 0.35018 \, \tau^{-1}$
         compared with the MD result for supercooled liquid Na at $T =
         216.3$ K.  From \cite{wall8}.}
\label{216.3Kvsmodel}
\end{figure}

\begin{figure}
\includegraphics{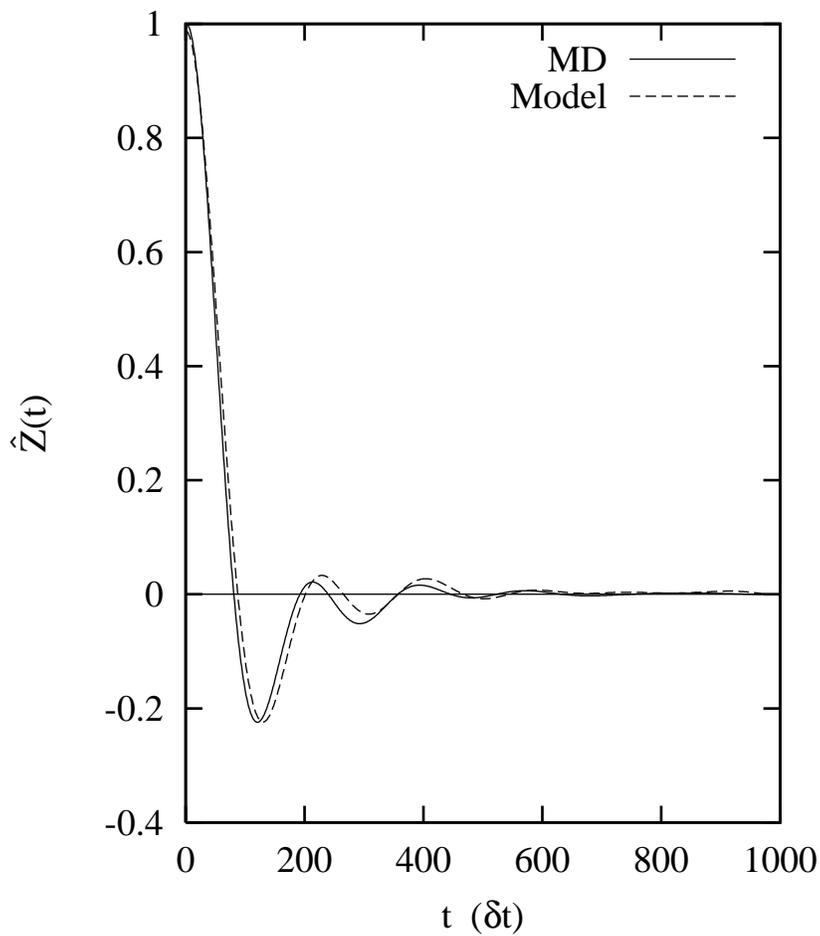}
\caption{The model prediction for $\hat{Z}(t)$ at $\nu = 0.83985 \, \tau^{-1}$
         compared with the MD result for liquid Na at $T = 425.0$ K.  From 
         \cite{wall8}.}
\label{425.0Kvsmodel}
\end{figure}

\begin{figure}
\includegraphics{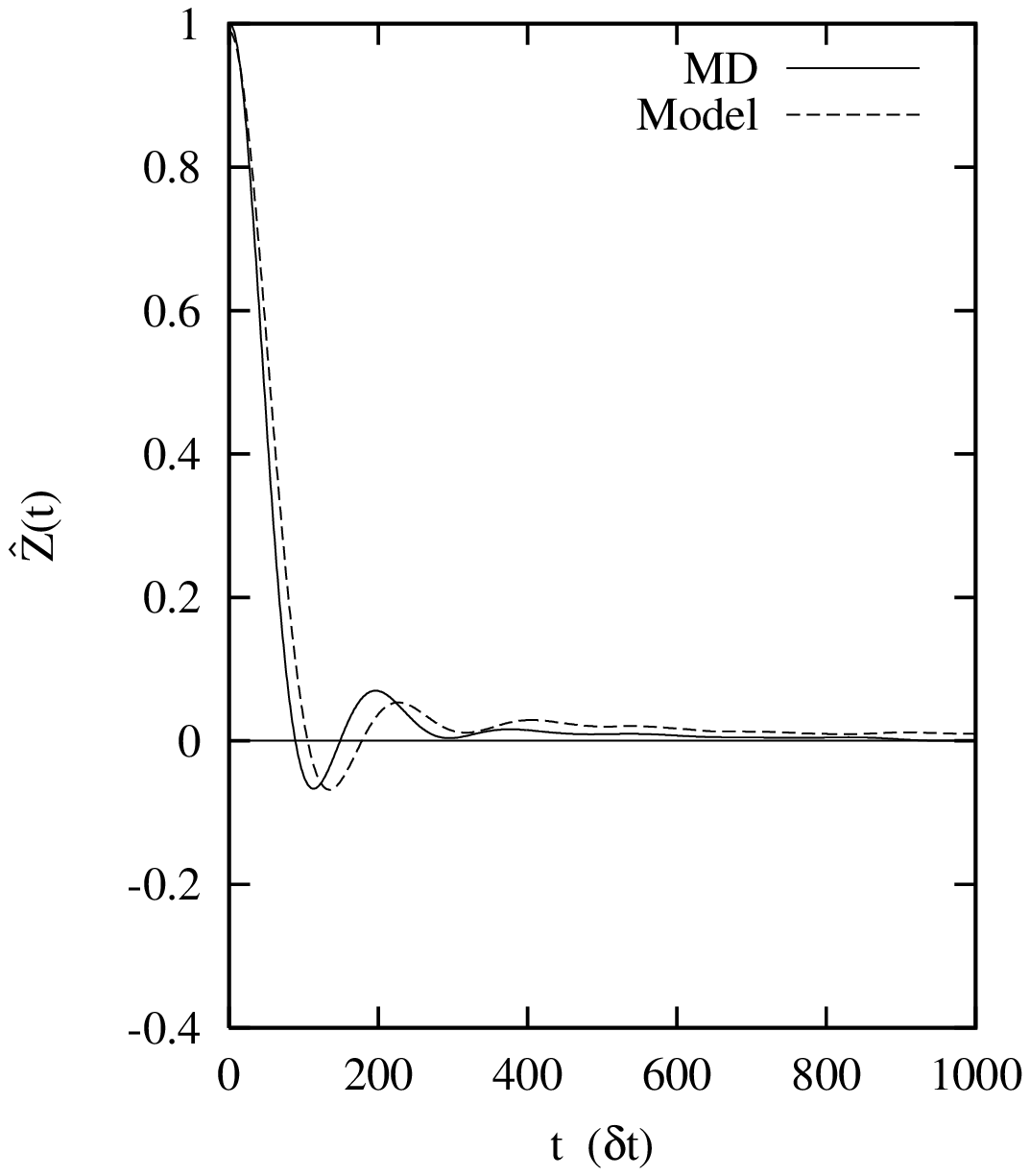}
\caption{The model prediction for $\hat{Z}(t)$ at $\nu = 1.68774 \, \tau^{-1}$
         compared with the MD result for liquid Na at $T = 1022.0$ K.  From 
         \cite{wall8}.}
\label{1022.0Kvsmodel}
\end{figure}

In \cite{wall8} we calculated $\hat{Z}(t)$ in this fashion and
compared the results with MD simulations of sodium with $N=500$ and
$\delta t = 0.2\,t^*$.  We performed \mbox{equilibrium} runs of the
system at temperatures ranging from the glassy regime to nearly three
times the melting temperature of 371 K\@.  At the two lowest
temperatures for which we ran MD, the system remained in a single
potential valley, as could be seen from examining the mean-squared
displacement; so these runs were compared to the model using $\nu =
0$.  For each of the higher temperatures, we ran the model for various
values of $\nu$, adjusting until the model matched the value of the
first minimum of $\hat{Z}(t)$.  Figures \ref{MDvsmodel} through
\ref{1022.0Kvsmodel} compare the model's predictions with a
representative sample of our MD results; the full set of results may
be found in \cite{wall8}.  In all figures, the transit rate is
expressed as a multiple of $\tau^{-1}$, where $\tau$, the mean
vibrational period in a random valley, is given in Subsection
\ref{transits}.  Note that all transit rates are on the order of
$\tau^{-1}$, supporting the contention made above that transits occur
roughly once per mean vibrational period.

\begin{figure}
\includegraphics{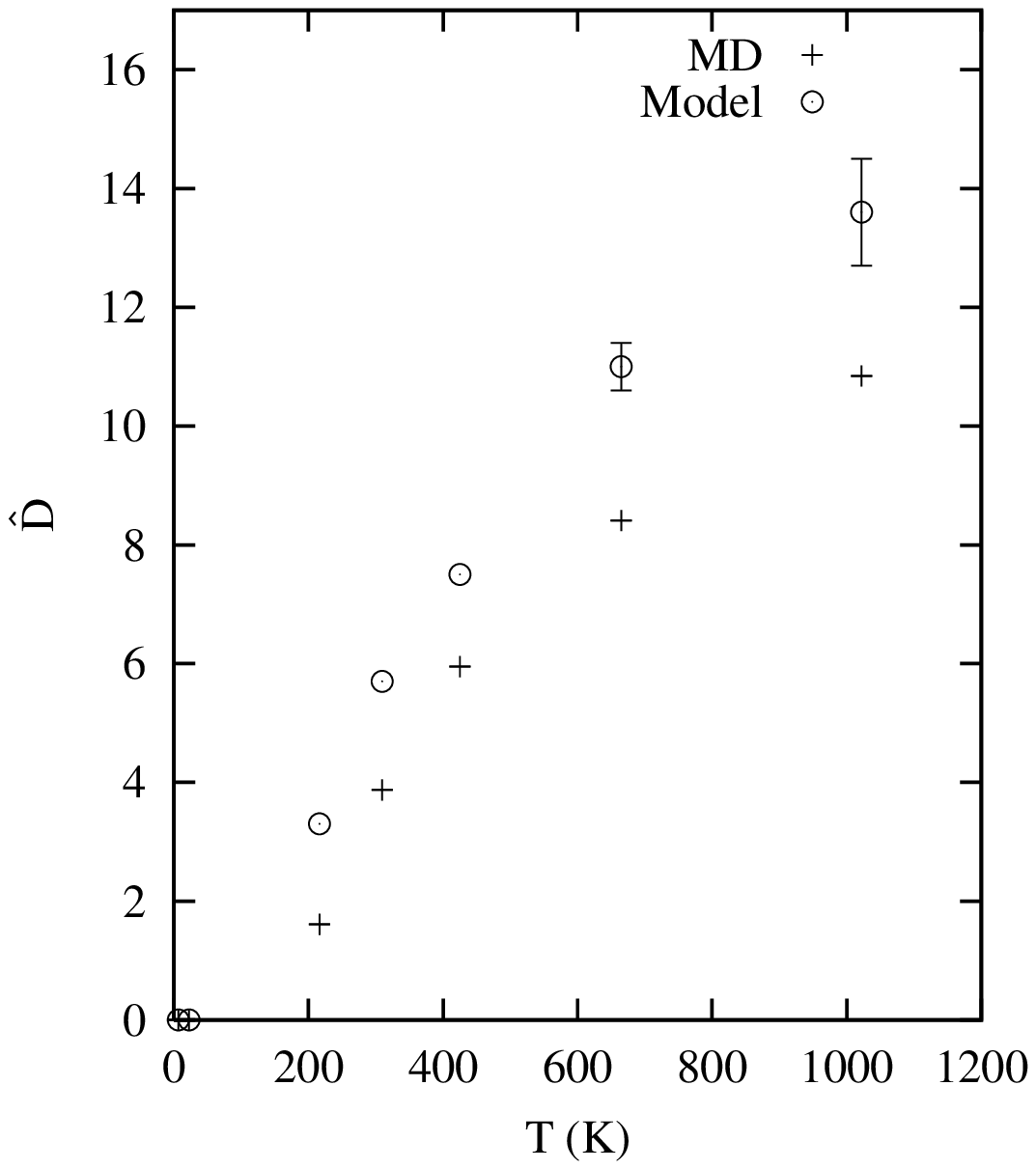}
\caption{$\hat{D}$ as a function of $T$ for both the model and MD.  From 
         \cite{wall8}.}
\label{DvsT}
\end{figure}

The most obvious trend in $\hat{Z}(t)$ is that its first minimum is
rising with increasing $T$; this is the primary reason for the
increasing diffusion coefficient $D$.  Note that the model is able to
reproduce this most important feature quite satisfactorily.  In fact,
all fits of the model to the MD results capture their essential
features, but we do see systematic trends in the discrepancies.
First, note that the location of the first minimum barely changes at
all in the model as $\nu$ is raised, but in MD the first minimum moves
steadily to earlier times as the temperature rises.  The first minimum
occurs at a time roughly equal to half of the mean vibrational period,
so the steady drift backward suggests that the MD system is sampling a
higher range of frequencies at higher $T$.  Also, in Figures
\ref{216.3Kvsmodel} and \ref{425.0Kvsmodel} the model tends to
overshoot the MD result in the vicinity of the first two maxima after
the origin, and in Figure \ref{1022.0Kvsmodel} this overshoot is
accompanied by a positive tail that is slightly higher than the (still
somewhat long) tail predicted by MD.  These overshoots should clearly
affect the diffusion coefficient $D$.  To check this, we calculated
the reduced diffusion coefficient $\hat{D}$, the integral of
$\hat{Z}(t)$, which is related to $D$ by $D = (k_B T/M) \hat{D}$.  The
results are compared to the values of $\hat{D}$ calculated from the MD
runs in Figure \ref{DvsT}.  This figure includes all of the data from 
\cite{wall8}, including the data not reproduced here. 
In all of the transiting cases, the model overestimates $\hat{D}$ by
roughly the same amount, which we take to be the effect of the
overshoots at the first two maxima.  At the higher temperatures the
discrepancy is also slightly higher, presumably due to the model's
long tail.

Although this single-particle model is promising, it is clearly based
on some arbitrary choices; possible improvements, taking advantage of
the information about transits from Subsection \ref{transits}, will be
discussed briefly in the next section.

\section{Outlook}
\label{outlook}

\subsection{What we've learned}

Two sets of experimental data on monatomic liquids, their specific
heats at the melting point and entropy of melting, led to two
hypotheses concerning their behavior:
\begin{enumerate}
\item The many-body potential surface of a monatomic liquid is composed of 
      approximately $w^N$ intersecting nearly-harmonic valleys which fall 
      into three classes: crystalline, symmetric, and random.  The random 
      class dominates the potential surface, and in the large-$N$ limit 
      these valleys all have the same depth, vibrational spectrum, and
      radial and angular distribution functions at the valley minimum.
\item The motion of the system decomposes into two types:  Oscillation in a 
      single many-body valley, and transits, which are nearly-instantaneous 
      transitions between valleys.    
\end{enumerate}
The picture that arises from these hypotheses has been tested
successfully with computer simulations of sodium and Lennard-Jones
argon, and it has been used to develop accounts of equilibrium and
nonequilibrium statistical mechanics of monatomic liquids which
compare very favorably with experiments and simulations.  Both of
these insights have demonstrated their value, and they should be taken
into account in any attempt at a comprehensive theory of monatomic
liquids.

\subsection{Further developments}

Work in any of the following areas would be of great interest.

\subsubsection{Studies of the potential energy landscape}

The crystalline valleys have been studied for decades and are by now
well understood.  Anharmonic effects in these valleys are complicated
but small in magnitude.  We need to know more, however, about the
random valleys, because of the dominant role they play in equilibrium
statistical mechanics.  The particle configurations of the random
valley structures in sodium, for example, need to be characterized
more completely than simply determining $G_{\gamma}(r)$.  Do they lack
any remnants of crystal symmetry, as asserted?  It would also be
worthwhile to continue the studies of argon above its critical density
(and other noble gas liquids) until its properties are as well
characterized as sodium's are.  The remaining nearly-free-electron
metals (22 or so elements) are expected to behave as sodium does,
considering the results of pseudopotential theory for these metals;
but that should be verified.  Finally, most of the remaining elements
in the periodic table that form monatomic liquids are
non-nearly-free-electron metals (such as the transition metals), and
electronic structure theory is just now reaching the point that their
interatomic potentials can be calculated reliably.  Whether they also
admit a division of their many-body potential valleys into similar
classes would be interesting to discover.

The symmetric valleys are the least studied in any of the elements we
have considered, and they will be important if we wish to understand a
real monatomic liquid that happens to quench into such a valley.  (An
experimental example of such a case is the amorphous carbon structure
in \cite{mcken}, cited in Subsection \ref{details}.)  The
distributions of $\Phi_0$ values and normal mode spectra $g(\omega)$,
among other quantities, should be determined.

Finally, we have asserted that the number of valleys is universally
$w^N$, where $\ln w = 0.8$, and that the randoms so outnumber the
others that virtually all the valleys are random.  Is this so?  Can
the valleys be counted?  It would be of tremendous interest to see if
the number of valleys approximately obeys this relation, since it is
crucial to much of the theory.

\subsubsection{Properties of anomalous melters}

Although the anomalous melters undergo substantial changes in their
electronic structure upon melting, they should obey the same liquid
dynamics theory as the normal melters; how they become liquids should
not affect how they behave once they are liquids.  However, as we saw
in Subsection \ref{exp}, the values of $\Theta_0$ and $S_E$ for these
elements should differ greatly between the liquid and crystal, and
these differences should account for the bulk of their entropy of
melting; testing this would be a very strong check on the theory of
liquids we have proposed.

\subsubsection{Extensions of equilibrium theory}

We have noted that all equilibrium thermodynamic quantities have both
anharmonic and boundary corrections, and theories of both of these
need to be developed.  Consider as an example $C_I$, the ionic part of
the specific heat, which the theory predicts to be $C_I = 3Nk_B +
C_{AB}$ (cf.\ Equation (\ref{classC})).  At the melting point, $C_I$
for the liquid, as for the crystal, shows small anharmonic effects,
and these appear to be of roughly the same sign and magnitude for both
phases (see Figure \ref{CV}).  Typically the full $C_V$ decreases as
$T$ increases, with $C_I$ ultimately falling to the value for the gas,
$(3/2)Nk_B$, and given the above comments on the anharmonic effects,
we expect the boundary correction to be mostly responsible for this
decrease.  One's classical intuition suggests that the boundary
correction is in fact negative, and this intuition is confirmed by
calculations of the correction resulting from cutting off the
potential of a one-dimensional harmonic valley \cite{wall2}.  Further
work on these corrections, however, remains to be done.

Another significant anharmonic effect which is not yet understood is
the fact that the equilibrium states occupying the random valleys in
Figure \ref{pvsk} and Figure 4 of \cite{wall4} do not follow a
straight line of unit slope at higher temperatures; this is why the
change in $\Phi_0$ between crystal and liquid is not closer to $T_m
\Delta S$, as discussed in Subsection \ref{exp}.  Is this feature
associated with the onset of diffusion?  Is it present in other
elements, or is it unique to sodium?  This effect is quite significant
and demands further study.

\subsubsection{Extensions of nonequilibrium theory}

We have discussed only one correlation function of interest, $Z(t)$,
and it remains to apply the picture to several others, such as the
stress-stress autocorrelation functions, which determine the shear and
bulk viscosities, and the dynamic structure factor $S(\mbox{\boldmath
$q$}, \omega)$, which determines the liquid's neutron scattering cross
section in the Born approximation.  Another area of interest is the
glass transition.  It has been shown that thermal properties of a
material during the glass transition depend on the cooling rate
\cite{ediger, angell1, vollmayr} and that if cooling or heating stops
while the system is undergoing the transition, it will then relax to
an equilibrium state \cite{angell1, angell2}.  This indicates that the
glass transition involves significant nonequilibrium behavior.  A
first attempt at a description of the glass transition using transits
may be found in \cite{wall7}, and further development of that line of
work is needed.

Several questions involving the picture's conception of transits also
need to be addressed.  First, does the picture accurately portray the
\mbox{mechanism} by which liquids diffuse at higher temperatures?  We
have seen that in low-temperature simulations transits occur in
precisely the manner predicted (Subsection \ref{transits}), but that
does not rule out the possibility of a qualitative change in behavior
as temperature increases.  What is the role of precursors and
postcursors, which currently are not incorporated into the picture?
Perhaps they indicate that the instantaneous transit is only a first
approximation, to be replaced by a more detailed process that unfolds
over a very small but finite time.  If the picture of transits needs
to be revised, then the revisions should affect the nonequilibrium
theory noticeably.  Then there is the specific transit model used in
our calculations of $Z(t)$: It accounts for the upward shift in the
first minimum in $Z(t)$, but it requires the transit amplitude to vary
as $T^{1/2}$ (because it equates the size of a transit to the
amplitude of oscillation of a typical particle), and a softer $T$
dependence is likely more accurate.  Also, the transit amplitudes it
predicts at the temperatures of the simulations in Subsection
\ref{transits} are smaller than the observed amplitudes by roughly a
factor of two.  Then, as we have already noted, in principle one
should be able to compute the transit rate $\nu$ using the Golden Rule
and the matrix element of the transition term in the Hamiltonian
between two states isolated in distinct valleys.  This would be a very
challenging calculation, but it would give us tremendous insight into
the mechanics of the transit process.  Note also that $\nu$ and the
boundary corrections $X_B$ to the thermodynamic quantities ultimately
arise from the same source: the boundary term in the Hamiltonian.  As
such, the two should be related, and a theory of that relationship
remains to be developed.

\subsection{The role of this theory}

This theory of monatomic liquid dynamics is based on a Hamiltonian,
from which both equilibrium and nonequilibrium properties follow, in
either quantum or classical regimes, according to the well-developed
principles of many-body physics.  The nearly harmonic character and
the statistical dominance of the random potential valleys render
equilibrium statistical mechanics tractable to leading order, and
they lead to well-defined corrections beyond leading order.
Decomposition of the motion into intra-valley oscillations and
inter-valley transits provides a basis from which time-correlation
functions can in principle be calculated from their definitions in
terms of the mechanical motion of the system.  In comparison, to our
knowledge QNM and INM theories have been developed only for the
calculation of correlation functions.  Both work with an averaged
normal mode frequency distribution $\langle g(\omega) \rangle$: QNM
theories average over configurations at the bottoms of potential
valleys, while INM theories compute a temperature-dependent average
over the entire configuration \mbox{space}.  Although neither of these
quantities enters the system's Hamiltonian, we can see that the QNM
$\langle g(\omega) \rangle$ can in principle approximate $g(\omega)$
for a single random valley.  When all is said and done, however, the
ultimate theoretical approach to this or any other problem is through
its Hamiltonian.  We believe that the ideas presented here provide a
useful framework for thinking about monatomic liquid dynamics, whether
one is refining equilibrium calculations to achieve improved agreement
with experiment or designing and analyzing experiments to learn more
about nonequilibrium processes.

\begin{center} \bf \Large Acknowledgments \end{center}

The authors gratefully acknowledge Brad Clements for his support and
collaborations.  This work was supported by the U.~S. DOE through
contract W-7405-ENG-36.

\end{document}